\documentclass{svjour2}                    
\usepackage{graphicx}
\usepackage{amssymb}
\usepackage{comment}
\usepackage{epstopdf}
\usepackage{color}
\usepackage{amsmath}
\usepackage{xcolor}
\usepackage{authblk}
\usepackage{pstricks}
\usepackage{tikz}
\usetikzlibrary{positioning}
\colorlet{linkequation}{blue}
\usepackage[colorlinks]{hyperref}

\usepackage[top=3cm, bottom=2cm,left=3cm, right=3cm,headheight=1.4cm,headsep=1cm]{geometry}

\newcommand{\bea}{\begin{eqnarray}}
\newcommand{\eea}{\end{eqnarray}}
\newcommand{\beq}{\begin{equation}}
\newcommand{\eeq}{\end{equation}}

\usepackage{hyperref}

\begin{document}

\title{On a ``continuum" formulation of the Ising model partition function
}


\author{Francesco Caravelli        
}


\institute{
Theoretical Division and Center for Nonlinear Studies,\\
Los Alamos National Laboratory, Los Alamos, New Mexico 87545, USA\\
\email{caravelli@lanl.gov}
}

\date{Received: date / Accepted: date}

\maketitle

\begin{abstract}
We derive an exact path integral formulation for the partition function for the Ising model using a mapping between spins and poles of a Laurent expansion for a field on the complex plane. The advantage in using this formulation for the evaluation of the partition function and $n-$point functions are twofold. First of all, we show that this mapping is independent of the couplings, and that for $h=0$ it is possible to perform a low temperature expansion as a perturbation theory via Feynman diagrams. The couplings are mapped naturally to a propagator for a complex field. The combinatorial nature of the partition function is shown to lead to an auxiliary field with a non-zero external field interaction which enforces the spin-like nature. Feynman diagrams are shown to coincide with certain combinations of traces of coupling inverses in a certain rescaling.
\end{abstract}

\section{Introduction}
The question we ask ourselves in this paper is how to construct a field theory for the Ising model for arbitrary exchange interaction couplings, and without having to resort to a continuous limit. The continuum formulation of the partition function we provide has to be thought as a compact way of keeping track of indices contractions in a field theory of the Ising model, and as a mere mathematical curiosity on its own. 

The Ising model is one of the oldest and most important models in Statistical Mechanics \cite{kittel,parisi,kadanoff,mussardo,baxter}. 
The model is simple enough to have a large range of applicability and it can be solved exactly in one, two dimensions and infinite dimensions. Various real systems can be mapped to Ising models, including models of the gas-liquid transition \cite{baxter}. In dimensions larger than four, the universality class is the same provided by the mean-field approximation. For the case in which there is a well established dimension and the couplings are all ferromagnetic, it is possible to derive a field theory as an expansion on the mean field. This is known as the Landau-Ginzburg model, and has been incredibly useful to study the order of phase transitions and via renormalization group, order dependent mapping \cite{zj} or variational perturbation theory \cite{fk}. 

The starting point of this articles is the fact that a field theory for the Ising model is not known for arbitrary couplings, as far as we know. The key difficulty is the fact that the continuous limit of a discrete operator does not always converge to a well-known continuous counterpart as in the case of the discrete Laplacian on a regular lattice of $D$-dimensions. The present paper proposes one formalism to overcome this difficulty, in part by hiding the discrete structure within a continuous operator, but without loosing the discrete nature of the model. 
This paper focuses on the formalism. In the next section we derive, with the use of the z-Transform, the mapping between the thermal field theory of the Ising model and an interacting field theory on the complex plane where the temperature is the perturbative parameter.   The mathematical identity we use is simply based on Cauchy's theorem for simple poles, and is a (rather simple) way of compactly performing indices contraction via (contour) integrations on the complex plane. 
We encode the spin-like nature of the Ising field in an auxiliary field via the introduction of of Dirac's deltas in the path integration. As we will see, the auxiliary field does not have a quadratic term, and thus does not allow a perturbative expansion. In order to overcome this difficulty, we introduce a quadratic term by hand, and consider a particular rescaling of the mass with the temperature in which we can obtain the partition function via the resummation only of certain type of diagrams, in a Hartree-Fock fashion and for zero external field. In this spirit, the resummation is in the spirit of a double-scaling limit. 

The interactions are then subsequently introduced, and peculiarity of the evaluation of the Feynman diagrams in this formalism described. The cases we study are in fact to elucidate the combinatorial properties of this field theory.
We show that a certain choice of the countour integration can be thought of as a Group Field Theory \cite{Oriti}.

\section{Mapping on the complex plane}
The Ising model has, despite its variety and richness in behavior, a simple Hamiltonian definition:
$$H=\sum_{ij} J_{ij} \sigma_i \sigma_j+\sum_i h_i \sigma_i,$$
where the variables  $\sigma_i=\pm 1$. Because of this, we can add an infinite constant to the Hamiltonian that can be reabsorbed into the diagonal $J_{ii}=m_i$, which contributes with $H_d=\sum_i m_i \sigma_i^2=\sum_i m_i$.
It is generally thought, however that it is not possible to write a consistent continuum theory when $J_{pt}$ do not have some special symmetry, as for instance Euclidean, but in the following no assumption is made about the nature of the couplings $J_{ij}$.

\subsection{Ising Hamiltonian as a continuous functional}
The main technique we use is a rather simple integral transform commonly used to solve difference equations and for signal processing. 
We define the following transformation, called $z$-Transform \cite{zt}, which maps countable variables to a function on the complex plane:
\begin{eqnarray}
\chi(z)&=&\sum_{i=1}^n \sigma_j z^{-j} \nonumber \\
\sigma_j&=& \frac{1}{2 \pi i} \oint_{\mathcal C} dz \chi(z)z^{j-1}=\frac{1}{2 \pi i} \oint_{\mathcal C} \frac{dz}{z } \chi(z)z^{j}
\label{eq:ztran}
\end{eqnarray}
where we assume that the contour $\mathcal C$ contains the origin in the complex plane, and $\chi(z):\mathbb C\rightarrow \mathbb C$. We see that the mapping above is based on Cauchy's integral theorem. The function $\chi(z)$ is not a smooth function in principle, as we do not require $\sigma_j$ to fall off in the index $j$ in any reasonable fashion. Since we aim to use this transformation to obtain a path-integral over any possible (non-differentiable) function $\chi(z)$, we do not see this as an obstruction. Some restriction will have however to be imposed on the coefficients $\sigma$ later on. We can however see that eqn. (\ref{eq:ztran}) provides the right answer as the poles of n-th order behave as Kronecker delta functions,
assuming that $\mathcal C$ is a contour that contains the origin in the complex plane.
These formulae can be generalized by moving the origin and thus the contour somewhere else in the complex plane, but in what follows this generalization will be necessary later as we will have to explore.
Let us now  use  this transformation on the Ising model.  A quick calculation shows that the partition function becomes
\begin{eqnarray}
Z&=&\sum_{\{\sigma\}} e^{-\beta \sum_{ij} \sigma_i \sigma_j J_{ij}} \nonumber \\
&=&\sum_{\{\sigma\}} e^{ \frac{1}{4 \pi^2 }\beta \oint_{\mathcal C}\oint_{\mathcal C} dz_1 dz_2 \chi(z_1)\chi(z_2) \sum_{ij} J_{ij} z_1^{i-1} z_2^{j-1}}.
\label{eq:hamint}
\end{eqnarray}
We note that $\chi(\frac{1}{z})$ is a Taylor polynomial whose coefficients are the spin variables.
The advantage of this transformation is that the summation can now be hidden in a continuous function,  $J(z_1^{-1},z_2^{-1})\equiv\sum_{ij} J_{ij} z_1^{i-1} z_2^{j-1}$, which is close to the definition of 2-dimensional z-Transform where the variables have been inverted. It is not too difficult to show that the mapping provided by the z-Transform is a very compact way of performing matrix multiplication for discrete systems (e.g. where we can replace the index contraction with a complex integration). 

 In fact, it can be written as:
\begin{equation}
(z_1z_2)^{-1}\sum_{ij} J_{ij} z_1^{i} z_2^{j} \equiv (z_1 z_2)^{-1} J(z_1^{-1},z_2^{-1})
\end{equation}
where $J(z_1,z_2)=\sum_{ij} J_{ij} z_1^{-i} z_2^{-j}$. Thus, given a certain integration measure which we now describe, we now show that we can write the partition function as a functional integration over the field $\chi$:
\begin{eqnarray}
Z&=& \int [D\chi(z)] P(\chi) e^{ \frac{\beta}{4 \pi^2 } \oint_{\mathcal C}\oint_{\mathcal C} \frac{dz_1 dz_2}{z_1 z_2} \chi(z_1)\chi(z_2)  J(z_1^{-1},z_2^{-1})}.\nonumber \\
\end{eqnarray}
The external field term $\sum_i h_i \sigma_i= \frac{1}{2 \pi i} \oint_{\mathcal C} dz \chi(z) \sum_i h_i z^{i-1}$ becomes
\begin{equation}
\sum_i h_i \sigma_i=\frac{1}{2\pi i}\oint_{\mathcal C} \frac{dz}{z} \chi(z) h(z^{-1}). 
\label{eq:scalar}
\end{equation}
Let us show that the functional we have written in eqn. (\ref{eq:hamint}) is correct, and poles do not interfere with each other. We have:
\begin{eqnarray}
H&=&-\frac{1}{(2\pi)^2} \oint_{\mathcal C}  \oint_{\mathcal C} dz_1 dz_2 \chi(z_1)\chi(z_2) z_1^{-1} z_2^{-1} J(z_1^{-1},z_2^{-1}) \nonumber \\
&=&-\frac{1}{(2\pi)^2} \sum_{ij} \sum_{qt} \oint_{\mathcal C}  \oint_{\mathcal C} dz_1 dz_2 \sigma_i \sigma_j  J_{qt} z_1^{q-i-1} z_2^{t-j-1}  \nonumber \\
&=&-\frac{1}{(2\pi)^2} \sum_{ij} \sum_{qt}  \sigma_i \sigma_j J_{qt} \oint_{\mathcal C}  dz_1  z_1^{-i+q-1} \oint_{\mathcal C}dz_2 z_2^{-j+t-1}  \nonumber \\
&=&-\frac{1}{(2\pi)^2} \sum_{ij} \sum_{qt}  \sigma_i \sigma_j J_{qt} \left(-(2\pi)^2\right) \delta_{i,q}\delta_{j,t} \nonumber \\
&=& \sum_{ij} \sigma_i \sigma_j J_{ij}.
\end{eqnarray}
The method can also be generalized to the case with interaction or n-body terms, which will be discussed elsewhere.

\subsection{Scalar product and operator action}
Before we discuss the path integral measure, it is worth explaining in a slightly more detailed fashion how the integral transform can be interpreted.
It is obvious that the integral in eqn. (\ref{eq:scalar}) defines a scalar product between vectors, which is inherited from the fact that the functions were originally scalar products. We thus define
\begin{equation}
\langle f(z), g(z)\rangle=\frac{1}{2\pi i}\oint_{\mathcal C} \frac{dz}{z} f(z^{-1}) g(z).
\label{eq:scalarprod}
\end{equation}
Albeit not obvious from the integral formula, the scalar product above is symmetric. The scalar product above will not be surprising for those familiar with Calogero-Sutherland-Moser models and Jack polynomials \cite{csm}.

The integral operation of eqn. (\ref{eq:scalarprod}) can be extended also to the action of operators, which are defined as functions $Q:\mathbb C\times \mathbb C\rightarrow\mathbb C$. 
Let us write $Q(z_1,z_2)=\sum_{ij} Q_{ij} z^{i} z^{j}$. Given two operators $Q_1(z_1,z_2)$ and $Q_2(z_1,z_2)$, we define
\begin{equation}
(Q_1 \circ Q_2)(z_1,z_2)=\frac{1}{2\pi i}\oint_{\mathcal C} dz\ z^{-1} Q_1(z_1,z^{-1}) Q_2(z,z_2)
\end{equation}
which is easy to see it being is non-commutative. In fact, we have
\begin{eqnarray}
(Q_1 &\circ& Q_2)(z_1,z_2)=\frac{1}{2\pi i} \oint_{\mathcal C} Q_1(z_1, z^{-1}) Q_2( z, z_2)  z^{-1} d z  \nonumber \\%
&=&\frac{1}{2\pi i} \sum_{ij,kt}\oint_{\mathcal C} (Q_1)_{ij} (Q_2)_{kt} z_1^i  z^j   z^{-k},  z_2^t  z^{-1} d z \nonumber \\
&=&\frac{1}{2\pi i} \sum_{ij,kt}\oint_{\mathcal C} (Q_1)_{ij} (Q_2)_{kt} z_1^i  z^{j-k+1}  z_2^t  d z \nonumber \\
&=& \sum_{ij,kt}Q_{ij} Q_{kt} z_1^i \delta_{j,k}  z_2^t   = \sum_{it} \left(\sum_k (Q_1)_{ik} (Q_2)_{kt}\right) z_1^i   z_2^t   \nonumber \\
&\equiv&  (Q_1Q_2)(z_1,z_2),
\end{eqnarray}
from which we see that the operation inherits its non-commutativity from the standard matrix product.
It does make sense to define for what follows the operation
\begin{eqnarray}
 [Q_1\circ \cdots \circ Q_n](z_1, z_2)&=&\frac{1}{(2\pi i)^n} \oint_{\mathcal C}\cdots  \oint_{\mathcal C} Q_1(z_1,\tilde z_1)  \cdots Q_{n-1}(\tilde z_n^{-1}, z_2)\prod_{i=1}^{n-1} \frac{d\tilde z_i}{\tilde z_i}
\end{eqnarray}
defined upon the matrix given by the matrix product $Q=Q_1 Q_2 Q_3 \cdots Q_{n-1}$. This operation will turn to be useful later.
The operation $\circ$ can be extended also to the action of the linear operator on a complex function, as
\begin{equation}
(J\circ \chi)(z)=\frac{1}{2\pi i} \oint_{\mathcal C} \frac{d\tilde z}{\ \tilde z }J(z,\tilde z^{-1})\chi(\tilde z).
\end{equation}
Using the definition of the scalar product and the action of the linear operator on a function, the quadratic part of the Hamiltonian can be simply written as:
\begin{equation}
H_{\chi \chi}=\beta \langle\chi(z), (J\circ \chi)(z)\rangle.
\end{equation}
Thus, the Green function for the field $\chi$ is simply provided by $J^{-1}(z_1,z_2)$ defined using the inverse matrix $J_{ij}^{-1}$ if it exists, and is defined via the Cauchy integral operation. This result  will be useful in perturbation theory as it defines the propagator.
\subsection{A comment on the path integral measure}
What is important is the measure of the path integral over the field $\chi(z)$. First, we note that the sum over discrete spins can be written as $\sum_{\sigma_i}=2 \int d\sigma_i \delta(\sigma_i^2-1)$ as it is well known. From the continuous variables $\sigma_i$, we can derive the measure above directly from the definition of the partition function (up to an irrelevant constant). In fact, we have
$d \chi_i\equiv d \chi(z_i)=\sum_j d \sigma_j z_i^{-j}$,
from which it can be shown that, up to a Vandermonde determinant which is independent from the fields and which factors out, we can replace $d\sigma_i$ with $[D\chi(z)]$ up to a constant factor which depends on the embedding into the complex plane, but not on the fields. Let show the statement above with a little more care.

We distinguish the functional versus the normal delta functions $\Delta$ and $\delta$ respectively, but the connection between the two should be obvious.
But we are interested in writing the integration over $[d \sigma]$ as $[d\chi(z)]$. This can be done via the introduction of a Jacobian, via the transformation:
from which we obtain the Jacobian determinant:
\begin{equation}
[d \chi(z_i)]= |\text{det}\left(\frac{\partial \chi(z)}{\partial \sigma_j }\right)| \prod_id\sigma_j.
\end{equation}
We have that $\frac{\partial \chi(z_i)}{\partial \sigma_j}=z_i^{-j}$ and thus it follows that the Jacobian is a Vandermonde determinant, given by $M=\text{det}(\frac{\partial \chi(z_i)}{\partial \sigma_j})=\prod_{p\neq q} (z_p-z_q)$. While we have
\begin{equation}
[d\sigma]=M^{-1}(z) [d \chi(z)],
\end{equation}
it is clear that the determinant function can be however reabsorbed as a field redefinition.\footnote{Another way to see this, is by noticing that 
we now use the following integration over Grassmanian variables $\sigma_i$ for the determinant of the matrix $M_{ij}=\delta_{ij} z^j$:
\begin{equation}
\text{det}(M_{ij})=\int \prod_i [d\theta_i] e^{-\sum_i \theta_i \theta_i z^i}.
\end{equation}
The Grassmann field is however decoupled from the Hamiltonian reformulation.}
\subsection{Spin nature of the field}
Insofar we have not achieved much as we still need to take care of the delta functions which provide the sum with the correct measure for the Ising model.
We note that the summation in the partition function over the value of the spins in eqn. (\ref{eq:hamint}) can be written as a continuous integral as:
\begin{eqnarray}
\sum_{\sigma_i\pm 1}&=&\prod_{i}\int d\sigma_i [\delta(\sigma_i-1)+\delta(\sigma_i+1)] \nonumber \\
&=&2^n \prod_i \int d\sigma_i \delta(\sigma_i^2-1) \rightarrow \int [d \sigma]  \delta(\sigma^2(x)-1).
\end{eqnarray}

The standard Landau-Ginzburg approach is to replace $\delta(x)\rightarrow e^{-\gamma x^2}$, and consider the action for large values of $\gamma$. Albeit possible, here we consider instead an auxiliary variable $\eta_i$, and write $\delta(\sigma_j^2-1)\propto \int d \eta_j e^{i \eta_j (\sigma_j^2-1)}$, simply because this is exact. For consistency with what follows, we introduce the notion of Cauchy delta function,
$\delta_C(z-z_0)=\frac{1}{2\pi i}\frac{1}{z-z_0}$, such that $\oint_{\mathcal C} dz f(z)\delta_C(z-z_0)=f(z_0)$.

Using the z-Transform again, we obtain:
\begin{eqnarray}
-i\sum_k \eta_k
&=&- \oint_{\mathcal C} z^{-1} \eta(z)\delta_C(z-1) dz \nonumber \\
i \sum_{k} \eta_k \sigma_k \sigma_k&=&-\frac{i}{4\pi^2} \oint_{\mathcal C} \oint_{\mathcal C} \frac{dz_1 dz_2}{z_1 z_2} \eta(z_1^{-1} z_2^{-1}) \chi(z_1) \chi(z_2) \nonumber
\end{eqnarray}
with $\eta(z)=\sum_{k} \eta_k z^{-k}$. Thus, using the z-Transform above we identify the seemingly innocuous $-i \sum_k \eta_k$ as an external field interaction with the auxiliary field $\delta_C (z)$.
From the point of view of a perturbative expansion for the path integral, it is interesting to note that the field $\eta$ does not have any quadratic term, which is something we will have to deal with later. We thus introduce this regularization rather artificially in the Hamiltonian, as 
\begin{eqnarray}
H_{\eta\eta}&=&-\frac{m^2}{4\pi^2}\oint_{\mathcal C}\frac{\eta (z) \eta(z^{-1})}{2} z^{-1} dz,
\label{eq:m}
\end{eqnarray}
with the purpose of studying perturbations theory. A suitable way of doing the limit $m\rightarrow 0$ has, however, to be devised. 
If we consider all the terms together, we find the following partition function for the Ising model:
\begin{eqnarray}
Z&=&\int [D\chi(z)] [D\eta(z)]e^{-\oint_{\mathcal C} dz z^{-1} \eta(z) \delta(z-1)+ \frac{1}{4 \pi^2 } \oint_{\mathcal C}\oint_{\mathcal C} dz_1 dz_2 \chi(z_1)\chi(z_2)z_1 ^{-1} z_2^{-1}\left(-i\eta(z_1^{-1} z_2^{-1})+ \beta J(z_1^{-1},z_2^{-1})\right)}\nonumber \\
&\cdot&e^{-i\beta \frac{1}{2\pi } \oint_{\mathcal C} dz_1 z_1^{-1} \chi(z_1) h(z_1^{-1}) -\frac{m^2}{4\pi^2}\oint_{\mathcal C}\frac{\eta (z) \eta(z^{-1})}{2} z^{-1} dz},
\label{eq:fieldtheory}
\end{eqnarray}
in which we have two interacting fields $\eta$ and $\chi$.
The path integral above is one of the key results of this paper, and which we will now study. The path integral is equivalent to the Ising model in the limit $m\rightarrow 0$, but as mentioned we do not require to have any other scale going to zero. We also note that the Hamiltonian is quadratic in $\chi$. There is an interaction term $\chi\chi\eta$, together with the external field interaction of $\eta$, enforces the spin-like origin of the field $\chi$, which makes the model non-trivial. At a first sight, we might wonder why the Landau-Ginzburg Hamiltonian is the truncated, continuum version of an Ising model, meanwhile the Hamiltonian above is quadratic. Clearly, the quadratic nature of the energy described by eqn. (\ref{eq:fieldtheory}) is only an illusion, and the interaction with $\chi$ generates an infinite tower of effective vertices for the field $\eta$. In fact, interacting quadratic field theories, with cubic interactions and with a complex interaction constant are the intermediate field representation of a $\phi^4$ scalar field theory.
Also, since at the bare level the $\eta$ field does not technically propagate, it cannot be interpreted as a fractionalization of a $\phi^4$ theory unless we introduce a mass term. We will discuss this later in this paper. 
Also, if $\mathcal C$ is chosen to be the unit circle, we note that any polynomial function defined on the unit circle will satisfy the property $\chi(z^{-1})=\chi^*(z)$ if the coefficients are real. In this case, the partition function that we have introduced is exactly a particular Group Field Theory on the unitary group with a non-trivial propagator for the field $\chi(z)$ \cite{Oriti}. 

What can be evaluated using the partition function above are general correlators of the form
\begin{equation}
G_N(z_1,\cdots,z_N)=\langle \chi(z_1)\cdots \chi(z_N) \rangle,
\end{equation}
from which we can obtain information on the spins from an inverse z-Transform. Intermediate calculations can be done via various methods, perturbative or non-perturbative, via the accumulated knowledge obtained in field theory.

\subsection{Where are the dimensions?}
What might seem suspicious of the mapping above is that we have mapped any notion of dimensionality into a one dimensional one. Such impression is only superficial.
The notion of dimensionality is not necessarily connected to the dimension of the embedding of the field, but rather to the notion of locality of the quadratic form associated with the propagator. In this sense, we have traded low dimensionality with locality, and the z-Transform mapping is not necessarily a simplification but a reformulation that demands to be studied.  For certain systems, the infinite sum contained in the operator $J(z_1,z_2)$ can be done explicitly. In fact, in those cases where dimensionality is well known, the function $J(z_1,z_2)$ can be calculated exactly. Consider for instance the case of the one dimensional Ising model, $J_{ij}=J_0 (\delta_{i,j+1}+\delta_{i+1,j})$. The transformed polynomial is given by:
\begin{equation}
\tilde J(z_1,z_2)=\sum_{ij}^\infty \delta_{i,j+1} z_1^i z_2^j=-  \frac{J_0}{2} \frac{z_1+z_2}{1-z_1 z_2}
\end{equation}
Analogously, the field $\chi(z)\rightarrow \chi(\vec x)$ becomes 
\begin{equation}
\sigma_{i_1,\cdots,i_D}=\frac{1}{(2\pi i)^D}\oint_{\mathcal C}\cdots \oint_{\mathcal C}\chi(z_1,\cdots,z_D) \prod _{\nu=1}^D z_\nu^{i_\nu} dz_\nu
\end{equation}
with $$\chi(\vec z \in \mathbb{C}^D)=\sum_{i_1,\cdots,i_n} \sigma_{i_1,\cdots,i_n}   \prod_{k=1}^D z_k^{-i_k-1}.$$


Using the Fourier transform for lattices, we can obtain a formula for the matrix $J_{kl}$ and its inverse, as
\begin{eqnarray}
J_{kl}&=&\frac{1}{(2\pi)^D} \int_{\mathcal B} d^D p G(p) e^{i(\vec k-\vec l)\cdot \vec p} \nonumber \\
G(p)&=&\epsilon-2 J \sum_{\nu=1}^D \cos(p_\nu)
\label{eq:exactinv}
\end{eqnarray}
and thus obtain the exact inverse in the Fourier representation:
\begin{eqnarray}
J(\vec z_1,\vec z_2)&=& \int_{\mathcal B} \frac{d^D p}{(2\pi)^D} \frac{G(\vec p)}{ \prod_{\nu=1}^D \frac{(1-e^{i p_\nu}z_{1\nu} )}{e^{i p_{\nu}}z_{1\nu}}  \frac{(1-e^{-i p_\nu}z_{2\nu}) }{e^{-i p_\nu} z_{2\nu} }}  \nonumber \\
\label{eq:exactinv2}
\end{eqnarray}
where $\vec z_i=(z_{i1},\cdots,z_{id})$. Also, the formulae above provide an exact representation for the trace. Since $\text{Trace}\left(J(z_1,z_2)\right)=\frac{1}{2 \pi i} \oint_{\mathcal C} \frac{dz}{z} J(z,z^{-1})=\sum_{i} J_{ii}$, we have simply 
$$\text{Tr} \left(J^{-1}\right)=\frac{1}{(2\pi)^D} \int_{\mathcal B} d^D p\ G^{-1}(p), $$
as one would expect. We have checked that the cofficients were right by direct inspection of all the possible Wick contractions.

It is also easy to see that we can define the Fourier transform of a function as:
\begin{eqnarray}
\tilde \chi( p)&=&\frac{1}{2\pi i} \oint_{\mathcal C} \frac{dz}{ z^{-1}} \chi(z^{-1}) \frac{e^{i p} z}{e^{i p}z-1} \nonumber \\
\chi(z)&=&\frac{1}{2 \pi} \int_{-\frac{\pi}{a}}^\frac{\pi}{a} dp \frac{e^{-i p} z}{e^{-i p}z-1} \tilde \chi (p).
\end{eqnarray}
The transforms above can be easily be generalized to $D$ dimensions via the introduction of a z-Transform for every index of the field $\sigma_{i_1,\cdots, i_D}$.

\section{Some trivial cases}
Let us discuss now a couple of simple cases where we can show that the path integral does indeed provide a connection between spin models and interacting scalar fields on the complex plane.

The fact that the path integral seems to be so simple does not mean that the approach above is simpler than other exact methods. For instance, we could not recover insofar any known results on 1-dimensional or 2-dimensional ferromagnetic models. 
This is due to the fact that we replaced the spin-like nature of the system with a scalar Yukawa interaction, which generates an infinite tower of interactions. In part, this is due to the fact that if we could solve the approach above exactly, we would solve any Ising model. Here we discuss two trivial cases in this formalism.

\subsection{Case $m=0$, $J=0$, $h\neq 0$}
This case is rather trivial, but it will help in boosting the confidence that the field theory we introduce does correspond to a spin model.
In the case $J=0$, if the approach above is correct, we should be able to recover the standard $Z=\prod \cosh(\beta h)$.
This case is sufficiently non-trivial in the approach we propose that requires a section on its own.
Let us first work out the case of a single spin. This will enlighten the fact that, albeit with some care, the integral over $\eta$ and $\sigma$ can be inverted.
We have:
\begin{equation}
\sum_{\sigma\pm1} e^{-\beta \sigma h}=2 \cosh(\beta h)
\end{equation}
We write:
\begin{eqnarray}
\sum_{\sigma\pm1} e^{-\beta \sigma h}&=&2\int_{-\infty}^\infty d\sigma \delta(\sigma^2-1) e^{-\beta \sigma h} =\frac{1}{\pi}\int_{-\infty}^\infty d\sigma \int_{-\infty}^\infty d\eta\ e^{i(\sigma^2-1)\eta-\beta \sigma h}
\end{eqnarray}
We now invert the order of the integration:
\begin{eqnarray}
Z&=&\frac{1}{\pi}\int_{-\infty}^\infty d\sigma \int_{-\infty}^\infty d\eta\ e^{i(\sigma^2-1)\eta-\beta \sigma h} \nonumber \\
&=&\frac{1}{\pi} \int_{-\infty}^\infty d\eta\int_{-\infty}^\infty d\sigma\ e^{i(\sigma^2-1)\eta-\beta \sigma h} \nonumber \\
&=&\frac{1}{\pi} \int_{-\infty}^\infty d\eta \frac{\sqrt{\pi } e^{\frac{i \beta ^2 h^2}{4 \eta }-i\eta}}{\sqrt{-i \eta }}
\end{eqnarray}
We now perform the following Wick rotation: $\eta\rightarrow i \eta$, and obtain
\begin{equation}
Z=-2  \frac{\sqrt{\pi }}{2\pi i}\int_{-i\infty}^{i\infty} d\eta \frac{ e^{\eta+\frac{ \beta ^2 h^2}{4 \eta }}}{\sqrt{ \eta }}
\end{equation}
and, as it turns out, the integral above is a representation of the hyperbolic cosine function \cite{ris}:
$$\frac{\sqrt{\pi }}{2\pi i}\int_{\epsilon-i\infty}^{\epsilon+i\infty} d\eta \frac{ e^{\eta+\frac{ \beta ^2 h^2}{4 \eta }}}{\sqrt{ \eta }} =\cosh(\beta h),$$
for $\epsilon>0$. We thus recover the same result.

Let us now discuss the case with multiple non-interacting spins.
We use the integration we have performed before. We have
\begin{eqnarray}
Z&=&\int [D\chi(z)] [D\eta(z)]e^{-\oint_{\mathcal C} dz z^{-1} \eta(z) \delta(z-1)}
e^{- \frac{i}{4 \pi^2 } \oint_{\mathcal C}\oint_{\mathcal C} dz_1 dz_2 \chi(z_1)\chi(z_2)z_1 ^{-1} z_2^{-2} \eta(z_1^{-1} z_2^{-1})} \nonumber \\
&\cdot& e^{-i\beta \frac{1}{2\pi } \oint_{\mathcal C} dz_1 z_1^{-1} \chi(z_1) h(z_1^{-1}) } \nonumber \\
&=&\int [D\eta(z)] Q(\eta) e^{-\oint_{\mathcal C} dz z^{-1} \eta(z) \delta(z-1)} e^{i \beta^2 \oint_{\mathcal C}\oint_{\mathcal C} dz_1 dz_2z_1 ^{-1} z_2^{-1} \frac{h(z_1^{-1}) h(z_2^{-1})}{4}A(z_1,z_2)^{-1}}\nonumber 
\label{eq:fieldtheory2}
\end{eqnarray}
where $A(z_1^{-1} z_2^{-1})=\eta(z)^{-1}\delta(z-z_1^{-1} z_2^{-1})$ and $Q(\eta)=\frac{C}{\sqrt{(-i)^n \text{det}(\eta(z_1^{-1} z_2^{-1}))}}$ with $C$ an integration constant. We note that the (functional) inverse of a function $A(x,y)$ is such that
$\int dz A(x,z) A^{-1}(z,y)=\delta(x-y)$. In this sense, the Dirac delta function is its own inverse, as $\int dz \delta(x-z) \delta(z-y)=\delta(x-y)$. Thus, $A(z_1,z_2)^{-1}=\eta(z)^{-1}\delta(z-z_1^{-1} z_2^{-1})$. We now use the discretization of the function $\eta(z_1,z_2)=\eta(z_i,z_j)=\sum_k \eta_k e^{i(\theta_i+\theta_j)(k-1)}$. This is the product of three matrices. In fact we can write
\begin{equation}
\eta(z_k,z_t)=\sum_{pq} e^{i \theta_k p} \delta_{pq}\eta_p e^{i\theta_j q}=(M \text{diag}(\eta_i) M^t)_{kt}
\end{equation}
where $M_{kt}=e^{i \theta_k (t-1)}$. Thus $\text{det}(\eta(z_i,z_j))=\text{det}^2(M) \text{det}((diag(\eta))=\text{det}^2(M)\prod_i \eta_i$.
We choose a discretization of the z variables along the circle such that it matches that number of components in $\eta_i$, $\sigma_i$ and $h_i$, and take the joint continuum limit $N\rightarrow \infty$.
As such, also the inverse operator can be written as
\begin{equation}
\eta(z_k^{-1},z_t^{-1})^{-1}=\sum_r M^{-t}_{kr} \eta_r^{-1} M^{-1}_{rt}
\end{equation}
We now have that the matrix $M$ is a DFT transformation with $\omega=e^{i \frac{2\pi}{N}}$. Its inverse simply requires the change of sign at the exponent. Thus, in order to calculate the inverse, we show that
\begin{eqnarray}
Q_{k,r}&=&\sum_p e^{i \theta_k t} e^{-i \theta_t r}=\sum_{k,r}e^{i \frac{2\pi}{N}kt -i \frac{2\pi}{N}tr} \nonumber \\
&=&\sum_t e^{-i \frac{2\pi}{N} t(k-r)}=\begin{cases}
0 & \text{ if } k\neq r \\
N & \text{ if } k=r
\end{cases}
\end{eqnarray}
which is non-zero only if $k=r$.
We do not have to go through the discretization, however. In the continuum, the operator $\eta^{-1}(z_1^{-1},z_2^{-1})$ can be obtained from the ansatz $\eta^{-1}(z_1^{-1},z_2^{-1})=c\sum_{k_2} \eta_{k_2} ^{-1} e^{-i(\theta_1+\theta_2)k_2}$, for a certain constant $c$, which has to satisfy
\begin{eqnarray}
\langle\eta,\eta^{-1}\rangle&=&\oint dz_2 z_2^{-1} \eta(z_1^{-1},z_2^{-1}) \eta^{-1}(z_2^{-1},z_3^{-1}) \nonumber \\
&=&-i\int_0^{2\pi} d\theta_2 c\sum_{k_1,k_2} \frac{\eta_{k_1}}{\eta_{k_2}} e^{i(\theta_1+\theta_2)k_1-i(\theta_2+\theta_3) k_2} \nonumber \\
&=&\delta(\theta_1-\theta_2)
\end{eqnarray}
We now observe that $\int_0^{2\pi} d_{\theta_2} e^{i\theta_2(k_1-k_2)}=-2\pi i \delta_{k_1,k_2}$. From which we obtain 
\begin{eqnarray}
\oint dz_2\ z_2^{-1} \eta(z_1^{-1},z_2^{-1}) \eta^{-1}(z_2^{-1},z_3^{-1})&=& 2\pi \sum_{k} e^{i(\theta_1-\theta_3) k} \nonumber \\
&=&2 \pi \delta(\theta_1-\theta_3)
\end{eqnarray}
Thus, we can set $c=\frac{1}{2\pi}$.
 The term off-diagonal terms are zero as the elements are roots of unity. A quick calculation shows that the integration over $z_1$ and $z_2$ can be performed, and thus
\begin{eqnarray}
F(h,h)&=&\oint _{\mathcal C} \oint_{\mathcal C}  dz_1 dz_2 z_1^{-1} z_2^{-1} \frac{h(z_1^{-1}) h(z_2^{-1})}{4}A(z_1,z_2)^{-1} \nonumber \\
&=&\int_0^{2\pi} \int_0^{2\pi} d\theta_1 d\theta_2 \frac{h_i h_j}{\eta_k} e^{i \theta_1 (i-k)} e^{i \theta_1 (j-k)}
\end{eqnarray}
from which we evince that the only surviving term is the one wih $i=j=k$ and gives a factor
\begin{equation}
\oint _{\mathcal C} \oint_{\mathcal C}  dz_1 dz_2 z_1^{-1} z_2^{-1} \frac{h(z_1^{-1}) h(z_2^{-1})}{4}A(z_1,z_2)^{-1}= \sum_i \frac{h_i^2}{4\eta_i}
\end{equation}

We thus obtain, using the result on the single spin obtained in the previous section, by means of $\eta_k \rightarrow -i \eta_k$, that 
\begin{equation}
Z\propto\prod_i \cosh(\beta h_i),
\end{equation}
which is the same result up to an external field redefinition. This shows the correctness of the functional approach above.

\subsection{No constraints: the case without auxiliary field and generalization}
It is clear that the auxiliary field imposes the constraint on the path integral to sum over functions which have only terms in the Laurent expansion which take a finite number of values.
If $\sigma$'s are free to be chosen over the real line, then the partition function does not require an auxiliary field $\eta$ and the model is equivalent to the Gaussian one \cite{parisi}.
In this specific case, the partition function integration should read:
\begin{eqnarray}
Z&=&\int [D\chi(z)] e^{ \frac{\beta}{4 \pi^2 } \oint_{\mathcal C}\oint_{\mathcal C} dz_1 dz_2 \chi(z_1)\chi(z_2)z_1 ^{-1} z_2^{-2}  J(z_1^{-1},z_2^{-1})-i\beta \frac{1}{2\pi } \oint_{\mathcal C} dz_1 z_1^{-1} \chi(z_1) h(z_1^{-1}) } \nonumber \\
&\propto&\frac{1}{\sqrt{\text{det}(\tilde J(z_1,z_2))}}e^{ \frac{\xi}{2} \frac{\beta^2 }{ (i 2\pi)^2} \frac{4\pi^2}{\beta}\oint_{\mathcal C}\oint_{\mathcal C} dz_1 dz_2 z_1^{-1} z_2^{-2} h(z_1^{-1})h(z_2^{-1}) \tilde J^{-1}(z_1,z_2)}.
\label{eq:fieldtheoryfree}
\end{eqnarray}
The constant $\xi$ can be fixed by the knowledge of the exact integral in the starting case, as we can write
\begin{eqnarray}
Z(\vec h)&=&\int_{-\infty}^\infty d\sigma_i e^{-\beta \sum_{ij} J_{ij} \sigma_i \sigma_j-\beta \sum_i h_i \sigma_i} =\frac{\sqrt{2\pi }^n}{\sqrt{\text{det}(\beta J)}} e^{ \frac{\beta}{2} \sum_{ij} h_i h_j J_{ij}^{-1}}.
\end{eqnarray}
We thus need to convince ourselves that $\oint_{\mathcal C}\oint_{\mathcal C} dz_1 dz_2 z_1^{-1} z_2^{-2} h(z_1^{-1})h(z_2^{-1}) \tilde J^{-1}(z_1,z_2)\propto \sum_{ij} h_i h_j J_{ij}^{-1}$. Using the result on the inverse operator from the previous sections,  it is straightforward to note that we can  fix $\xi=\frac{i}{2\pi }$. We  have
\begin{eqnarray}
Z&=&\int [D\chi(z)] e^{ \frac{\beta}{4 \pi^2 } \oint_{\mathcal C}\oint_{\mathcal C} dz_1 dz_2 \chi(z_1)\chi(z_2)z_1 ^{-1} z_2^{-2}  J(z_1^{-1},z_2^{-1})} e^{-i\beta \frac{1}{2\pi } \oint_{\mathcal C} dz_1 z_1^{-1} \chi(z_1) h(z_1^{-1}) } \nonumber \\
&=&\frac{e^{  \frac{\beta}{4\pi i} \oint_{\mathcal C}\oint_{\mathcal C} dz_1 dz_2 z_1^{-1} z_2^{-2} h(z_1^{-1})h(z_2^{-1}) \tilde J^{-1}(z_1,z_2)}}{\sqrt{\text{det}(\tilde J(z_1,z_2))}}. \nonumber \\
\label{eq:fieldtheoryfree2}
\end{eqnarray}
From the analysis above, we can also see for instance that 
\begin{equation}
\frac{1}{\beta \delta h(z_1^{-1})}\frac{1}{\beta \delta h(z_2^{-1})} Z=\mathbb E [\chi(z_1) \chi(z_2) ]= \frac{1}{\beta} J^{-1}(z_1,z_2)
\end{equation}
from which we obtain the result that the continuous spin correlator.

\section{Interactions}

\subsection{Field rescaling for perturbation theory}
The path integral formulation we obtained above is completely equivalent to the partition function of the Ising model, and in fact it has to be thought not as a novelty, but as a compact way to keep track of the indices contractions. In fact, it goes without saying that one could introduce the integral for each spin, the auxiliary spins, and use this field theory, as it is usually done in thermal field theory. However, we have found some peculiarities of this formalism novel enough to write a paper on its own on the subject.

We now discuss how to use the field theory in the complex plane we discussed using perturbation theory. The perturbative coupling constant, which is the inverse temperature, is in fact on the propagator. In order to perform the perturbative expansion however, we need to move the coupling to the interaction term. Thus, we consider the rescaled fields $\chi\rightarrow \sqrt{\frac{1}{\beta}} \chi$, and we consider $h(z)=0$ for simplicity. Then, the propagator term for $\chi$ looses $\beta$, while the term $\chi \chi \eta$ acquires a coupling constant $g=\frac{T}{J}$ which we can use for a perturbation theory.

Here we make a comment about how to interpret a perturbative expansion of the interaction term $\chi \chi \eta$ at $h=0$. In principle the expansion is well defined for $\beta$ finite, and if the propagator $J(z_1,z_2)$ is positive. 
We can add a diagonal term to the interaction, equivalent to an infinite constant, $-nm^2=-m^2 \sum_i \sigma_i^2$, and use $m$ as a regularization of the integrals. The quadratic term thus becomes of the form $-m^2 \delta_{ij}+\beta J_{ij}\rightarrow -m^2 \frac{z_1 z_2}{(1-z_1 z_2)} + \beta J(z_1,z_2)$.
We simplify the notation below by omitting the integrals in the complex plane. We have
\begin{equation}
L=-\beta H+L_{\eta}=m^2 \chi \chi-\beta \chi \chi J+ \chi \chi \eta + \eta J_{\eta} - \beta \chi h.
\end{equation}
We can redefine $\chi \rightarrow \beta^{-\frac{1}{2}} \chi$ (assuming $\beta\neq 0$), from which the Lagrangian reads
\begin{equation}
L= -m^2 \eta \eta-\chi \chi J+ \beta^{-1} \chi \chi \eta + \eta \delta_{C} - \beta^{\frac{1}{2}}   \chi h.
\end{equation}

If $h=0$, we have that the expansion of the interaction term can be done as a low temperature expansion over the Gaussian theory. In fact, $\eta$ enforces the spin nature of the model, and naturally this gives a perturbative way to evaluate physical quantities for the Ising model. If we assume that this expansion is valid, we have at $h=0$:
\begin{eqnarray}
Z&=&\int[D\chi][D\eta] e^{\oint_{\mathcal C}\left(H_\eta+H_\chi\right)} \sum_{k=0}^\infty \frac{1}{k!} ( T \oint \chi \chi \eta + \oint \eta \delta_C)^k.\nonumber
\end{eqnarray}
which is a perturbative expansion with a measure which is a Gaussian model for $\chi$ and an external field and mass for $\eta$.


\subsection{Feynman rules}
We have stressed enough the importance of the interaction between the field $\chi$ and $\eta$ for the correct analysis of the model.
We define $J(z_1^{-1},z_2^{-1})=\sum_{ij} J_{ij} z_1^{i} z_2^j \equiv \tilde J(z_1,z_2)$, and where we have defined $J(z_2)\equiv h(z_2^{-1})$ and $K(z_2)\equiv \delta(z_2-1)$.
The partition function:
\begin{eqnarray}
Z&=&\int [D\chi(z)] [D\eta(z)]e^{-\oint_{\mathcal C} dz z^{-1}\left( \eta(z) \delta(z-1)-\frac{m^2}{4\pi^2} \eta(z) \eta(z^{-1})\right)}\nonumber \\
&\cdot&e^{- \frac{i}{4 \pi^2 } \oint_{\mathcal C}\oint_{\mathcal C} dz_1 dz_2 \chi(z_1)\chi(z_2)z_1 ^{-1} z_2^{-2} \left(-i \beta^{-1} \eta(z_1^{-1} z_2^{-1})+ \tilde J(z_1,z_2)\right)}\nonumber \\
&\cdot&e^{-i\sqrt{\beta} \frac{1}{2\pi } \oint_{\mathcal C} dz_1 z_1^{-1} \chi(z_1) h(z_1^{-1}) }
\end{eqnarray}
\begin{figure*}
\includegraphics[scale=0.3]{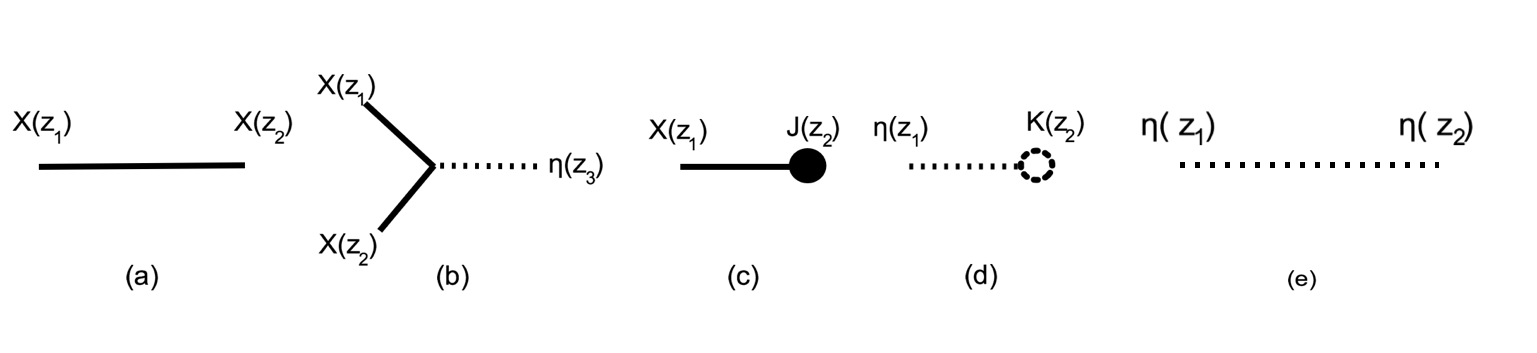}
\caption{Feynmann diagrams associated with the field theory of eqn. (\ref{eq:fieldtheory}). Also, each external line of $\eta(z)$ contributes with $m$, while each external line of $\chi(z)$ contributes $2\pi i$.}
\label{eq:feynmannrules}
\end{figure*}
We now discuss the field theory properties of the model above.
In Fig. \ref{eq:feynmannrules} we show the Feynman diagrams associated with eqn. (\ref{eq:fieldtheory}). For each point $z$, one associates an integral $\oint_{\mathcal C} dz\ z^{-1} \cdot$. 
Then, we have the following associated rules:
\begin{eqnarray}
\text{(a)} &=& 4\pi^2 \tilde J^{-1}(z_1,z_2) \nonumber \\
\text{(b)} &=& -\frac{i}{4 \pi^2 \beta}\delta(z_3-z_1^{-1} z_2^{-1}) \nonumber \\
\text{(c)} &=& -i\frac{\sqrt{\beta}}{2\pi} h(z_2^{-1}) \nonumber \\
\text{(d)} &=& -\delta(z_2-1) \nonumber \\
\text{(e)} &=& -4 \pi^2 m^{-2} \delta(z_1-z_2^{-1})
\end{eqnarray}

In order to see whether this is correct, we can integrate out the field $\chi$ explicitly, and write
the effective action as
\begin{eqnarray}
Z(h(z))&=&\int [d\eta] e^{-\oint_{\mathcal C} dz z^{-1}\left( \eta(z) \delta(z-1)-\frac{m^2}{4\pi^2} \eta(z) \eta(z^{-1})\right)+\frac{\beta^2}{2}\oint dz_1 dz_2 h(z_1^{-1}) h(z_2^{-1}) (\eta(z_1^{-1} z_2^{-1})}\nonumber \\
& &\ \ \ \ \  \cdot e^{\beta \tilde J(z_1,z_2))^{-1} z_1^{-1} z_2^{-1} - \frac{i}{2} \text{Tr}\log(\eta(z_1^{-1} z_2^{-1})+\beta \tilde J(z_1,z_2))} \nonumber
\end{eqnarray}
which can be the beginning of a Loop-Vertex-Expansion as we realize that the action was in an intermediate field representation  \cite{rivasseau}. This possibility will considered in future works \cite{inprep}.
\subsection{A few examples}
\begin{figure}
\centering
\includegraphics[scale=2]{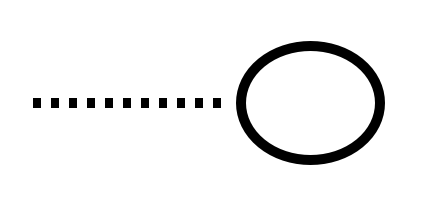}
\includegraphics[scale=2]{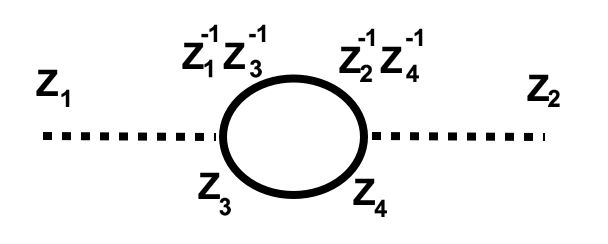}
\caption{External field and effective propagator for the $\eta$ field.}
\label{fig:effpropeta}
\end{figure}

Let us now look at the effective interaction at the first order  in $T$ for the field $\eta$.
As a first comment, we note that an effective external field is provided in Fig. \ref{fig:effpropeta} (top), which takes the form
\begin{eqnarray}
J^1_{\eta, eff}(z_1)&=&-\frac{i\sqrt{z_1} T }{2\pi} \oint_{\mathcal C} dz  z^{-1} \tilde J(\sqrt{z_1} z, \sqrt{z_1}^{-1}z^{-1}) \nonumber \\
&=& \sqrt{z_1} \ \text{Trace}(J^{-1}).
\end{eqnarray}
We thus obtain that 
\begin{equation}
    J_{\eta,eff}=\delta_C(z-1)+\sqrt{z} T \sqrt{Z} Q
\end{equation}
with $Q=\text{Trace}(J^{-1})$, and depends on the traces of the Green function. This is a feature of diagrams with only external $\eta$ legs, and we see that it comes naturally in this formalism.

\begin{figure*}
\centering
\includegraphics[scale=1.5]{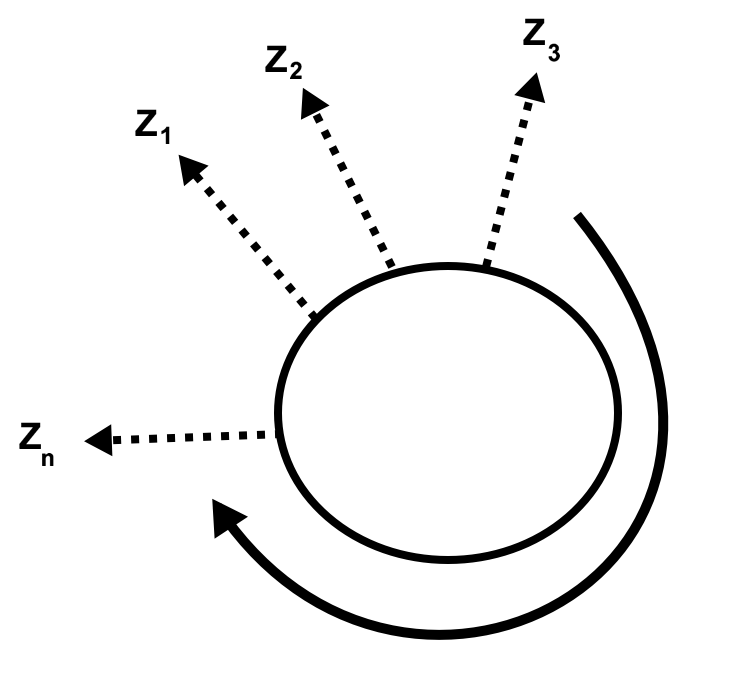}\includegraphics[scale=1.5]{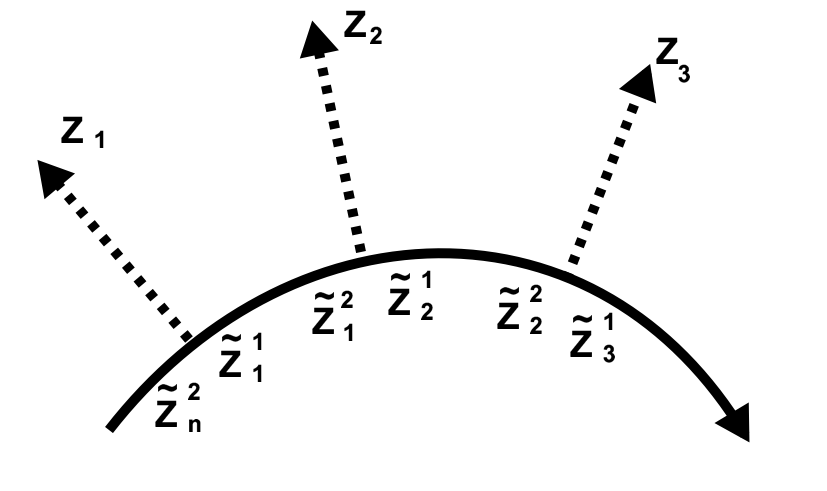}
\caption{The interaction vertex $n$-point function for $\eta(z)^n$ and its parametrization. We call these \textit{urchin diagrams}.}
\label{fig:nvert}
\end{figure*}

It is interesting to note that we can also introduce the effective propagator for the $\eta$ field, mediated via the $\chi$ interaction.
This is shown in Fig. \ref{fig:effpropeta} (right) and is given by
\begin{eqnarray}
G(z_1,z_2)
&=& -\frac{(4\pi^2)^2 \sqrt{z_1 z_2}}{\beta^2}  \text{Trace}(J^{-2})
\end{eqnarray}
and we note that the term $V=\sum_{ij} (J_{ij}^{-2})=\text{Trace}(J^{-2})$, which is the sum of all the elements of the matrix $J_{ij}$ squared is, in principle, divergent if not properly regularized. 
Since in what follows these diagrams are important, 
let us now consider the $n-$point function  as in Fig. \ref{fig:nvert}. We call these \textit{urchin diagrams} and will be relevant later when we perform a double scaling limit.
The diagram can be written as:
\begin{equation}
V_n(z_1,\cdots,z_n)=D_n \frac{(4\pi^2)^{2n}}{\beta^n} \frac{(-i)^n}{ (4 \pi^2)^n}\oint_{\mathcal C} \left(
 \prod_{i=1}^n d\tilde z^1_i d\tilde z^2_i (z_i^1)^{-1}(z_i^2)^{-1} \tilde J^{-1}(\tilde z^1_i,\tilde z^2_i)\right) \delta(z_1- \tilde z_{n}^2 \tilde z_{1}^1)\prod_{k=1}^{n-1} \delta(z_k- \tilde z_{k}^2 \tilde z_{k+1}^1)
\end{equation}
where $D_n$ is a symmetry factor on which we will focus on later.  We can rewrite the integral as
\begin{eqnarray*}
V_n(z_1,\cdots,z_n)&=&D_n (-i)^n \frac{(4\pi^2)^n}{\beta^n}\nonumber \\
&\cdot& \prod_{i=1}^n \sqrt{z_i} \oint_{\mathcal C} 
  d\tilde z_i  \tilde J^{-1}(\sqrt{z_n}\tilde z_n, \sqrt{z_{1}}^{-1}\tilde z_1^{-1}) \nonumber \\
  &\cdot&\prod_{i=1}^{n-1}\tilde J^{-1}(\sqrt{z_i}\tilde z_i, \sqrt{z_{i+1}}^{-1}\tilde z_i^{-1}) \nonumber \\
  &=&D_n (-i)^n \frac{(4\pi^2)^n}{\beta^n} \prod_{i=1}^n \sqrt{z_i} (2 \pi i)^n \text{Trace}(J^{-n}) \nonumber \\
  &=&D_n \frac{(2\pi)^{3n}\text{Trace}(J^{-n})}{\beta^n  }  \prod_{i=1}^n \sqrt{z_i}   \nonumber \\  
\end{eqnarray*}
We thus see that these diagrams are naturally associated with traces of powers of the Green function. We also now understand the role of the external field  $\delta_C$ for the $\eta$ auxiliary field in the combinatorics.
For a symmetric tridiagonal matrix with zero elements on the diagonal and elements $J$ of the upper and lower diagonals, the eigenvalues are given by $2J \cos(\frac{k \pi}{(N+1) })$ for $k=1\cdots N$ and $N$ is the number of spins. 

\section{Resummation of the low temperature expansion in a mass parameter rescaling}
In this section we provide further background to the low temperature expansion, via constraining ourself to a very specific limit of the coupling constant, the temperature. Low temperature expansions for arbitrary couplings must necessarily be ill-defined. This is due to the Dyson argument for which, if the series expansion at $T=0$ was convergent, then we the series would also be defined for $T<0$, which from the point of view of Statistical Physics does not make sense. There are several known ways to avoid this situation: the first and most well known is to resort to renormalization, whether constructive, perturbative or non-perturbative. The second and simpler is to consider a subset of diagrams via coupling rescaling, and thus focus only on a certain phase of the model. We discuss the latter first.
\subsection{Temperature rescaling}
\begin{figure}
\centering
\includegraphics[scale=1.2]{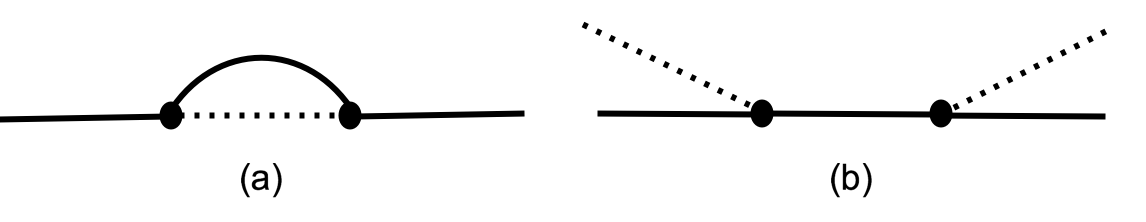}
\caption{Two diagrams contributing at order $T$. Diagram $(a)$ is of order $\frac{T^2}{m}$, while diagram $(b)$ is of order $\frac{T^2}{m^2}$. }
\label{fig:lowTexp}
\end{figure}
Let us first discuss one possible well-defined way to perform the limit $m\rightarrow 0$ diagrammatically.
Each  line of $\eta$ (both internal and external) contributes a factor of $m$.  For instance, in Fig. \ref{fig:lowTexp} we observe two diagrams which contribute at the order $T^2$. In the first, before rescaling, diagram $(a)$ gives $\frac{T^2}{m^2}$, meanwhile diagram $(b)$ is $\frac{T^2}{m^4}$. 
This is a general feature of the model: since for each interaction vertex there is a factor of $T$ associated, we can rescale $T\rightarrow T m^2$, and the limit $m\rightarrow0$ implies that, at the order $T^{k}$, only diagrams with $k$ external $\eta$ lines dominate. The limit $m\rightarrow 0$ then corresponds to the weak coupling regime which we are interested in.
After the rescaling aforementioned, we see  that the diagram (a) in Fig. \ref{fig:lowTexp} goes to zero in the limit $m\rightarrow 0$. This implies, again, that for each internal lines of $\eta$, at the same order of $T$, the diagram acquires a factor $m$ after the rescaling of the temperature.
This allows us to study only what we call \textit{urchin diagrams}, e.g. diagrams with only external $\eta$ lines, and is reminishent of a Hartree-Fock approximatino. Using the device of the limit $m\rightarrow 0$, we can also make sense of a low temperature expansion. We thus obtain that the limit:
\begin{equation}
\lim_{T,m\rightarrow 0} G^{T,m}(z_1,\cdots,z_n)\text{  :  }\frac{T}{m^2}=\text{constant}  
\end{equation}
is equivalent to the urchin diagrams expansion for the partition function that we study below. We thus keep the ratio $\frac{T}{m^2}=\tilde T$ fixed. The expansion above is similar in spirit to the melonic expansion in a certain class of group field theories and in constructive field theory, and is thus equivalent to keeping diagrams with the higher superficial degree of infrared divergences. Alternatively, we can also consider the strong coupling case, in which $T m$ is constant in the limit $m\rightarrow 0$. In this case, the external field interactions are set to zero, while diagram (a) in Fig. \ref{fig:lowTexp} effectively becomes the contraction of an effective propagator for a non quadratic theory (as the propagator for $\eta(z)$ is $\frac{1}{m}\delta(z-z^{-1})$): in this limit we reobtain Landau-Ginzburg interactions, and $\eta$ becomes an intermediate field representation for $\phi^4$ theory. The strong coupling regime will be discussed elsewhere.
\subsection{Combinatorics of closed loops, and set partitions}
Since we have argued that the double scaling limit above is a summation with only urchi diagrams, it is worth spending some time to explain the combinatorics of these diagrams. 

Given the urchin diagrams expansions above, we now discuss how to evaluate Feynman diagrams and the coefficient $D_n$ in a bit more detail. The evaluation of the Feynman diagrams takes advantage of few properties of the complex integrations we have mentioned before.
At the order $T^{k}$, we have $k$ vertices. Since the field $\eta$ does not propagate internally after the rescaling, the only possibility is that these appear as external fields. For the partition function at $h=0$, this implies for $Z$ that only diagrams with amputated legs of $\eta$ will appear. 
We define $G_n^c(z_1^1,z_1^2\cdots,z_n^1,z_n^2)$ the connected diagrams with $k$ vertices and $2n$ internal $\chi$ legs. 
For each external $\eta$ leg, there is a factor of $\delta(z-1)$. Since the external legs necessarily connect only to two internal legs, and if all internal legs are contracted, we have locally terms of the form
\begin{eqnarray}
\frac{1}{2\pi i}\oint_{\mathcal C} &d[z]& dz_\eta dz_{e^\prime}^1 z_{e^\prime}^2 z_{e}^1 dz_{e}^2 Q_{\mathcal G\backslash\{e,e^\prime\}} ([z])  \nonumber \\
&\cdot&J^{-1}(z_{e}^1,z_{e}^2) J^{-1}(z_{e^\prime}^1,z_{e^\prime}^2)  \nonumber \\
&\cdot&\delta(z_\eta-(z_e^2)^{-1}(z_{e^\prime}^1)^{-1}\delta(z_\eta-1)
\end{eqnarray}
where $d[z]$ contain integration over the rest of the internal legs and $z_e^1,z_{e^\prime}^2$, and $Q_{\mathcal G\backslash\{e,e^\prime\}} ([z])$ represents the rest of the graph. We see immediately that we can integrate out the external field $\delta(z_\eta-1)$ and the vertex, and impose $z_e^{2}=(z_{e^\prime}^1)^{-1}$.  This is the equivalent of the scalar product between the Green functions:
$$\frac{1}{2\pi i}\oint_{\mathcal C}  dz_{e^\prime}^1 J^{-1}(z_{e}^1,(z_{e^\prime}^1)^{-1}) J^{-1}(z_{e^\prime}^1,z_{e^\prime}^2)=J^{-2}(z_e^{1},z_{e^\prime}^2). $$
This is all we need to evaluate amputated diagrams in the following. 

For the case of the partition function, we thus need to consider all possible closed loops, as there are no external $\chi$ amputated lines, while for the case of $n-point$ functions of $\chi$, there can be open lines. However, it is easy to see that for closed lines necessarily we must have, following the integrations:
\begin{equation}
\frac{1}{2\pi i}\oint_{\mathcal C}  dz_{e}^1 J^{-k}(z_{e}^1,(z_{e}^1)^{-1})=\text{Trace}\left(J^{-k}(z_e^{1},z_{e^\prime}^2)\right).
\end{equation}
Thus, for an operator $Q(z_1,z_2)$, the trace is defined as 
\begin{equation}
\text{Trace}(Q)=\frac{1}{2\pi i}\oint_{\mathcal C} \frac{dz_1 dz_2}{z_1 z_2}\delta(z_1-z_2^{-1})Q(z_1,z_2).
\end{equation}
The result above shows that for each closed loop in $\mathcal G$ we associate a trace over a power of the Green function which is the length of the loop. 
In principle this could help in evaluating the coefficients in the Feynman diagrams.

Let us now use this result in conjunction with an exact mapping between the Feynmann diagrams with the set partitions. First we note that the combinatorial factor associated with a loop of length $r$ is $2^{r-1} (r-1)!$ due to the various ways in which 2-legs of a graph can be joint to make a loop. This factor is due to the mechanics of Wick contractions and the automorphism group of a cycle.

\begin{figure}
\centering
\includegraphics[scale=1.3]{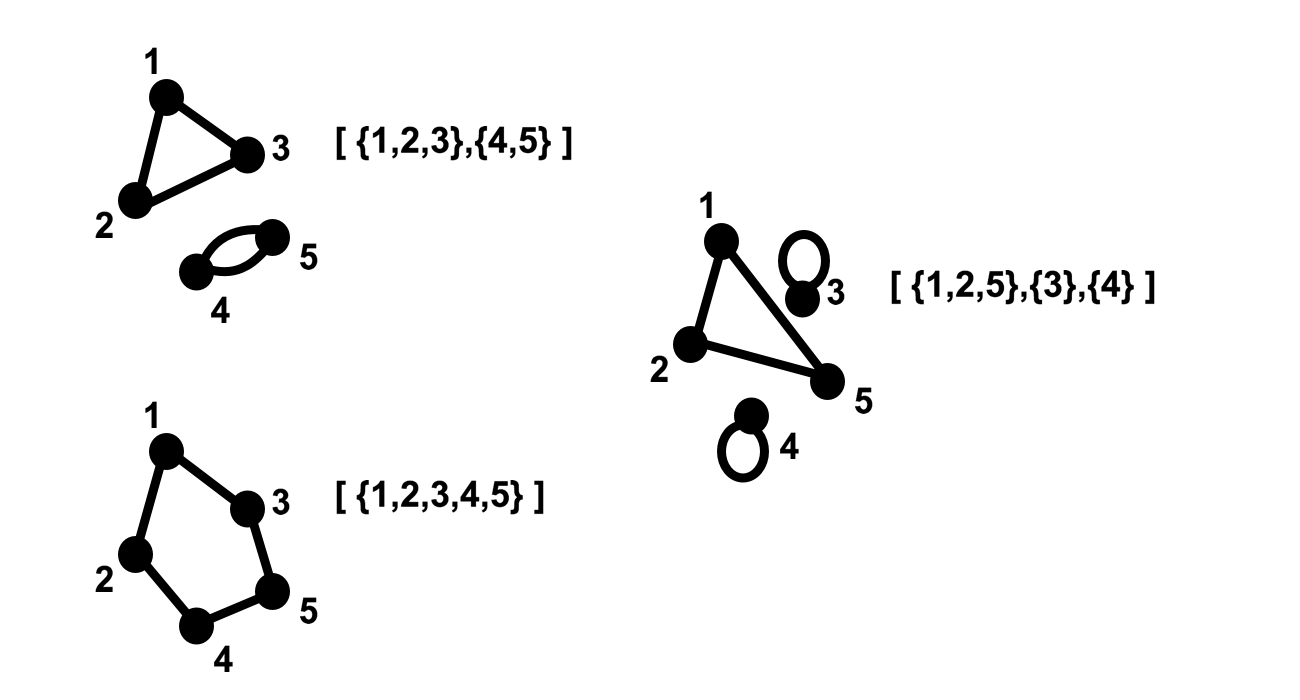}
\caption{Example of the mapping between graphs and the indices of integer partitions. Effectively, the combinatorics is the one of a loop gas.}
\label{fig:setpart}
\end{figure}

 In order to see the connection with the set partitions, let us now label the $n$-vertices of a graph as $v_1,\cdots, v_n$. It is easy to see that there is a one to one mapping to set partitions, as in Fig. \ref{fig:setpart}, and more specifically the partition of integers. For each cycle among nodes, we write the loop as a order-indendent partition of the nodes. To each partition between $n$ nodes, corresponds a certain value of the Feynman diagram. For instance, to the sets $[\{1,2,3\},\{4,5\}]$, $[\{1,2,3,4,5\}]$ and $[\{1,2,5\},\{3\},\{4\}]$, correspond $\text{Trace}\left(J^{-3}\right)\text{Trace}\left(J^{-2}\right)$, $\text{Trace}\left(J^{-5}\right)$ and $\text{Trace}\left(J^{-3}\right)\left(\text{Trace}\left(J^{-1}\right)\right)^2$ respectively, and obviously independently of the labelling of the nodes. We need, thus, to count identical partitions among the nodes in order to obtain the coefficient in front of a certain value of a Feynman diagram. We can thus write the sum over graphs $\mathcal G$ as a sum over set partitions, where the coefficients are the associated partitions of integers. Let $\pi \in  \Pi(k)$ be a certain partition of an integer $k$, $\{n_1,\cdots,n_p \}$, such that $\sum_i n_i=k$,  $n_i \in \mathbb N$ and $P_f$ the number of such identical partition of the set (here $n_i$ can be degenerate). Each loop of length $n_i$ will give a factor $\text{Trace}(J^{-n_i})$, but if there are many loops of identical length $g_i$, there will be a factor $\text{Trace}(J^{-n_i})^{g_i}$ in that diagram.

In order to calculate the combinatorial coefficient, we write the partition of the integer as a sum over the nondegenerate blocks \cite{toufik}:
$$ \sum_{i=1}^{|\pi|} g_i n_i=k$$
where $g_i$ is the number of partitions of identical size $n_i$, and $n_i$ are all distinct. For instance for a Feynman graph with 3-nodes, we will have
\begin{eqnarray}
\pi_1\rightarrow\ 3&=&1+1+1=3\cdot 1\nonumber \\
\pi_2\rightarrow\ 3&=&2+1\nonumber \\
\pi_3\rightarrow\ 3&=&3
\end{eqnarray}
which implies $|\Pi(3)|=3$, and we have $|\pi_1|=3, |\pi_2|=2, |\pi_3|=1$. 
The degeneracy of the size is thus encoded in the factor $g_i$. The formula above clearly classifies all partitions of integers. Then, we know that given $g_i$ and $n_i$, the total number of such partitions has a formula which is $\frac{k!}{g_i! (n_i!)^{g_i}}$. For $\pi_1$, we have only one block of degeneracy $g_i=3$, for $\pi_2$ we have two blocks both of degeneracy $g_i=1$, and in $\pi_3$ we have one block of degeneracy 1.

We thus have the formula
\begin{equation}
O(T^k):\ \  k!\sum_{\pi \in \Pi(k)}  \prod_{B_i\in \pi}^{|\pi|} \frac{\left((n_i-1)!2^{(n_i-1)}\right)^{g_i}}{g_i!( n_i!)^{g_i}}  (\text{Tr}\left( J^{-n_i} \right))^{g_i}.
\end{equation}
For $k=2$, there are only 3 contractions, of which 2 gives identical terms. For $k=3$ we have 15 contractions, meanwhile for $k=4$ we have 105 contractions. We will use the formula above in the following in order to check whether the coefficients are correct.  For $k=1$, the only diagram is the tadpole which gives at the order $O(T)$, $\text{Trace}(J^{-1})$. The combinatorial coefficients become already non-trivial at the subsequent order, where we have two partitions of the integer 2, $2=1+1$ ($g_i=2$, $n_i=1$) and $2=2$ ($g_i=1$, $n_i=2$):
\begin{eqnarray}
O(\tilde T^2)&:&\ \ \frac{1}{2!}\Big(\frac{2^{1(2-1)} 2!}{1! (2!)^1}\ \text{Tr}(J^{-2})+\frac{2^{2(1-1)} 2!}{2! (1!)^2} \left(\text{Tr}(J^{-1})\right)^2 \Big) \nonumber \\
&=&\ \ \frac{1}{2!}\Big(2\ \text{Tr}(J^{-2})+ \left(\text{Tr}(J^{-1})\right)^2 \Big) \nonumber \\
\end{eqnarray}
For instance, let us consider $k=3$, $3=2+1=1+1+1$. There is one partition with one block $g_1=3$ $n_1=1$,  one partition with two blocks $g_1=1$, $n_1=2$, and $g_2=1$ $n_2=2$  and one partition with one block $g_1=1$ and $n_1=3$. Thus, we immediately obtain
\begin{eqnarray}
O(\tilde T^3):\ \ \frac{1}{3!}\Big(8\ \text{Tr}(J^{-3})&+&6\ \text{Tr}(J^{-2})\text{Tr}(J^{-1}) \nonumber \\
&+&\left(\text{Tr}(J^{-1})\right)^3 \Big).
\end{eqnarray}
At the order $O(\tilde T^4)$, we have 105 Isserlis-Wick possible contractions. We have five possible partitions, 
$4=1+1+1+1=2+2=3+1=2+1+1$.
We thus obtain from our formula,
\begin{eqnarray}
O(\tilde T^4):\ \ \frac{1}{4!}\Big(&48&\ \text{Tr}(J^{-4})+32\ \text{Tr}(J^{-3})\text{Tr}(J^{-1}) \nonumber \\
&+&12\left(\text{Tr}(J^{-2})\right)^2 +\left(\text{Tr}(J^{-1})\right)^4  \nonumber \\
&+&12\left(\text{Tr}(J^{-1})\right)^2\text{Tr}(J^{-2})\Big)
\end{eqnarray}
and the coefficients sum up to $105$ as one would expect.

\subsection{Resummation}
From the previous sections, we see that in the double-scaling limit we can thus write the partition function as
\begin{equation}
    Z=\sum_{k=0}^\infty T^k \sum_{\pi \in \Pi(k)}  \prod_{B_i\in \pi}^{|\pi|}\frac{1}{g_i!} \left( \frac{2^{n_i-1} \text{Tr}\left( J^{-n_i} \right)}{n_i}  \right)^{g_i}.
\end{equation}
The formula above is thus connected to the spectral properties of the coupling matrix $J$.
Let us now try to write this sum in a less combinatorial way. First, we note that we can write $T^k=T^{\sum_i i g_i}$. For each number $k$, we can write $\sum_{\Pi(k)} $ as follows :  
\begin{equation}
    \sum_{\pi\in \Pi(k)}\rightarrow \oint \frac{dz}{2\pi i} \sum_{g_1,\cdots,g_k=0}^\infty \frac{1}{z^{k+1-\sum_{i=1}^k i g_i}}.
    \label{eq:trick}
\end{equation}
Something that will however provide an analytical final formula is the following remark. It is interesting to note that we can write also
\begin{equation}
    \sum_{i=1}^{\tilde k} i g_i=k,
    \label{eq:const}
\end{equation}
with $g_i\in \mathbb{N}$ and $\tilde k\geq k$. We know however that for $i>k$, $g_i=0$. 
As a quick check of this fact, note the following. Consider $k=3$ and $\sum_{i=1}^3 i g_i=k$, for which we have only $3=1+1+1=1+2$. From the integral formula we have introduced, we obtain $\text{Res}(\frac{1}{z^4 \prod_{j=1}^{\tilde k} (1-z^j) },0)=3$, $\forall \tilde k\geq 3$, which is true as it can be promptly checked. This implies that we can write term by term an infinite sum over zeros in the expansion, as we will see in a moment.  

We can write the constrained sum using an integral over the complex plane, via the imposition of the constraint of eqn. (\ref{eq:const}) using eqn. (\ref{eq:trick}). We have then,
\begin{eqnarray}
   \sum_{\pi \in \Pi(k)}  \prod_{B_i\in \pi}^{|\pi|}\frac{1}{g_i!} \left( \frac{2^{n_i-1} \text{Tr}\left( (\frac{T}{J})^{-n_i} \right)}{n_i}  \right)^{g_i}&=& \oint \frac{dz}{2\pi i} \sum_{g_1,\cdots,g_{\tilde k}=0}^\infty \frac{1}{z^{k+1-\sum_{i=1}^{\tilde k} i g_i}}\prod_{i=1}^{\tilde k}\frac{1}{g_i!} \left( \frac{2^{i-1} \text{Tr}\left( (\frac{T}{J})^{-i}  \right)}{i}  \right)^{g_i}  \nonumber \\
&=&\oint \frac{dz}{2\pi i}  \frac{1}{z^{k+1}}\prod_{i=1}^{\tilde k} \sum_{g_i=0}^\infty \frac{1}{g_i!} \left( \frac{2^{i-1} z^{i} \text{Tr}\left( (\frac{T}{J})^{-i}  \right)}{i}  \right)^{g_i}  \nonumber \\
&=&\oint \frac{dz}{2\pi i}  \frac{1}{z^{k+1}} e^{\frac{1}{2}\sum_{i=1}^{\tilde k} \frac{2^{i} z^{i} \text{Tr}\left( (\frac{T}{J})^{-i}  \right)}{i} } \nonumber \\
&=&\oint \frac{dz}{2\pi i}  \frac{1}{z^{k+1}} e^{\frac{1}{2}\sum_{n=1}^N\sum_{i=1}^{\tilde k} \frac{2^{i} T^i z^{i}}{\lambda_n^i i} }, 
\end{eqnarray}
where we have identified $n_i=i$.
We now use the identity
\begin{equation}
    \sum_{i=1}^{\tilde k} \frac{x^i}{i}=x^{\tilde k+1} (-\Phi (x,1,\tilde k+1))-\log (1-x),
\end{equation}
where $\Phi (x,1,k+1)$ is the Lerch Zeta-function, defined as
\begin{equation}
    \Phi(x,s,z)=\sum_{j=0}^\infty \frac{x^j}{(j+z)^s}.
\end{equation}

We now reconsider the normalization that we had ignored insofar.
We then obtain the following representation for the partition function for finite $\tilde k$:
\begin{eqnarray}
    \mathcal Z_{\tilde k}(T)&=& Z_{\text{Gauss}}(T)\oint \frac{dz}{2\pi i}F_{\tilde k}(z,\lambda) e^{-\frac{1}{2} \sum_{n=1}^N \log(1-\frac{2 z T}{\lambda_n})} \nonumber \\
    &=&Z_{\text{Gauss}}(T)\oint \frac{dz}{2\pi i} \frac{F_{\tilde k}(z,\lambda)}{\prod_{i=1}^N \sqrt{1-\frac{2 zT}{\lambda_i}}}
\end{eqnarray}
with 
$$F_{\tilde k} (z,\lambda)=\sum_{k=0}^{\tilde k} \frac{1}{z^{ k+1}}  e^{-\frac{1}{2} \sum_{n=1}^N (\frac{2 z}{\lambda_n})^{\tilde k+1} \Phi (\frac{2 z}{\lambda_n},1,\tilde k+1) }.$$

We would like to obtain a better expression from the above, and what we get can be intended in two ways. 
First, we are interested in the asymptotic values of the sum in $F$, from which we can simply take
\begin{equation}
    F(z,\lambda)\approx \frac{1}{z}\sum_{k=0}^\infty \frac{1}{z^{k}}\approx \frac{1}{z-1}.
\end{equation}
Alternatively, we can consider the limit
\begin{equation}
    \lim_{\tilde k\rightarrow \infty} Z_{\tilde k}\implies \lim_{\tilde k\rightarrow \infty} \sum_{j=0}^{\tilde k} \frac{x^j}{j}=-\log(1-x),
\end{equation}
that as we as seen should not contribute to the integral. In both cases, we obtain the expression
\begin{equation}
    \mathcal Z(T)= Z_{\text{Gauss}}(T) \oint \frac{dz}{2\pi i}\frac{e^{-\frac{1}{2} \sum_{n=1}^N \log(1-\frac{2 z T}{\lambda_n})}}{(z-1)}. 
\end{equation}
We can rewrite this equation more succintly in terms of the spectral density for the matrix $J$, as 
\begin{equation}
    \mathcal Z(T)=Z_{\text{Gauss}}(T)\oint \frac{dz}{2\pi i}\frac{e^{-\frac{1}{2} \oint d\lambda \rho_{J}(\lambda)  \log(1-\frac{2 z T}{\lambda})}}{(z-1)},
    \label{eq:nonpz}
\end{equation}
which is our final expression, with
\begin{equation}
    \rho_J (\lambda) =\sum_{\lambda_i \in \Lambda(J)} \delta(\lambda-\lambda_i).
\end{equation}
We have thus rewritten the partition function of the Ising model in terms of its spectral representation only. It is worth mentioning that this representation is rather hard to interpret, as it involved branch cuts in the complex plane that are due to the position of the eigenvalues of $J$. However, we note that there is the emergence of a non-perturbative pole at $z=1$, and it would be tempting to enlarge the domain to enclose it. It is not clear whether this procedure is however correct.

Another way of treating the partition function is via the tricks used in the study of the f-Mayer expansion, or linked cluster expansion \cite{Huang}. We start from
\begin{equation}
    Z=\sum_{k=0}^\infty T^k \sum_{\pi \in \Pi(k)}  \prod_{B_i\in \pi}^{|\pi|}\frac{1}{g_i!} \left( \frac{2^{n_i-1} \text{Tr}\left( J^{-n_i} \right)}{n_i}  \right)^{g_i}.
\end{equation}
and note that we can write it as
\begin{eqnarray}
    Z&=&\sum_{g_1=0}^\infty\cdots \sum_{g_N=0}^\infty \prod_{i} \frac{1}{g_i!} \left( \frac{2^{i-1} \text{Tr}\left( J^{-i} \right)}{i} T^i \right)^{g_i}, \nonumber \\
    &=& e^{\sum_{i=1}^\infty \frac{2^{i-1} \text{Tr}\left( J^{-i} \right)}{i} T^i } \nonumber \\
    &=&e^{-\frac{1}{2} \int d\lambda \rho(\lambda)\log(1-\frac{2T}{\lambda})}.
\end{eqnarray}
It is interesting to note that the expression above is exactly the one we would have obtained from eqn. (\ref{eq:nonpz}) by calculating the residue in $z=1$ rather than $z=0$. In this sense, the non-perturbative residue is not near the perturbative one.

Now we can obtain that
\begin{eqnarray}
    F&=&-\kappa T \log \mathcal Z \nonumber \\
    &=&-\kappa T \log Z_{\text{Gauss}}(T)+ \frac{\kappa T}{2} \int d\lambda \rho(\lambda) \log(1-\frac{2T}{\lambda}) \nonumber 
\end{eqnarray}
where 
\begin{equation}
-\kappa T \log Z_{\text{Gauss}}(T)=\frac{\kappa T}{2} \log \det \beta J= \frac{\kappa T}{2} \int d\lambda \rho(\lambda) \log \frac{\beta \lambda}{2}
\end{equation}
from which we get 
\begin{equation}
F=-\kappa T \log \mathcal Z
= \frac{\kappa T}{2} \int d\lambda \rho(\lambda) \log \left(\frac{\beta \lambda}{2}-1\right)
\end{equation}

Let us now look at the case of a $D$-dimensional lattice of size $L$.
The eigenvalues, in this case, are known to be parametrized as
\begin{equation}
    \lambda(n_1,\cdots,n_D)=2a + 2b \sum_{i=1}^D \cos (\frac{2\pi}{L} n_i)
\end{equation}
for $n_1=0,\cdots, L-1$. We can thus write the integral above as (we take the continuous limit):
\begin{eqnarray}
F&=&-\kappa T \log \mathcal Z \nonumber \\
&=& \frac{\kappa T L^D}{2 (2\pi)^D} \int_0^{2\pi} dp^D \log \left(\beta (a+b \sum_{i=1}^D \cos p)  -1\right)\nonumber 
\end{eqnarray}
The free energy per spin and per unit of energy can be written as
\begin{eqnarray}
\frac{f}{\kappa T}&=&\frac{F}{L^D \kappa T} \nonumber \\
&=& \frac{1}{(2\pi)^D} \int_0^{2\pi} dp^D \log \left(\beta (a+b \sum_{i=1}^D \cos p)  -1\right)\nonumber 
\end{eqnarray}
and look at its internal energy per site (at zero magnetic field), given by
\begin{equation}
    u(T)=-T^2 \partial_T \frac{f}{\kappa T}
\end{equation}
and the specific heat,
\begin{equation}
    c(T)= \partial_T u(T).
\end{equation}
We obtain the expression for the internal energy (we set $\kappa=1$):
\begin{equation}
    u(T)= \frac{1}{(2\pi)^D} \int_0^{2\pi} dp^D \frac{T(a+b \sum_{i=1}^D \cos p)}
    {\left( (a+b \sum_{i=1}^D \cos p)  -T\right)}
\end{equation}
We now use the standard trick of writing the inverse as
\begin{equation}
    \frac{1}{\lambda}=\int_0^\infty dt e^{-t \lambda} 
\end{equation}
and obtain (we assume that we are in an analytical region):
\begin{equation}
    u(T)= \int_0^\infty dt \frac{T}{(2\pi)^D} \int_0^{2\pi} dp^D (a+b \sum_{i=1}^D \cos p)
    e^{-t\left( (a+b \sum_{i=1}^D \cos p)  -T\right)}
\end{equation}
This term can be written in two parts:
\begin{eqnarray}
u(T)&=& a T \int_0^\infty dt\ e^{-t(a-T)} J_0(-ibt)^D \nonumber \\
&+& b D T \int_0^\infty dt\ e^{-t(a-T)} J_1(-i bt)^D.
\end{eqnarray}
Asymptotically ($t\gg 1$), we see that
\begin{equation}
    J_{\alpha}(-ib t)\approx \frac{1}{\sqrt{2 \pi bt}} (\cos(-i b t-\frac{ \pi}{4}) (2 \alpha +1))
\end{equation}
Thus for $b<0$ and $t\gg 1$ we obtain
\begin{eqnarray}
u(T)&=& 2(a+b D) T \int_0^\infty dt\  \frac{e^{-t(a+ b D-T)}}{(2 \pi b t)^\frac{D}{2}} .
\end{eqnarray}
which the internal energy as a function of $T$.
Thus, the resummation of Urchin diagrams is not completely trivial, e.g. we do not reobtain the Gaussian theory.

\section{Interactions with the external field}
Before we go into the final remarks, we would like to discuss the interactions when also the external field $h$ is included.
Let us consider the full model using the Green connected functions, defined as
\begin{equation}
G_{n,\text{conn}}^{n,k}(z_1,\cdots,z_n,\tilde z_1,\cdots,\tilde z_k)
\end{equation}
is all the connected diagrams which satisfy the Feynman rules. We now note that since the only interacting vertex has two $\chi$ lines and one $\eta$ lines, and the $\eta$ field does not propagate, the number interacting vertices equals to the number of outgoing $\eta$ lines.
We have that the generator of connected diagrams
\begin{eqnarray}
W\left(h_{\eta},h_{\chi} \right)&\equiv& \log\left(Z\left(J_{\eta},J_{\chi} \right)\right) \nonumber \\
&=&\sum_{n=0}^\infty \frac{1}{n!}   \sum_{k=0}^{\frac{n}{2}} \oint_{\mathcal C} G^{n,2k}_{n,conn}(z_1,\cdots,z_n,\tilde z_1,\cdots,\tilde z_{2k}) \nonumber \\
&\cdot& \prod_{i=1}^n h_{\eta}(z_i) z_i^{-1} dz_i  \prod_{r=1}^k \beta h_{\chi}(\tilde z_r) \tilde z_r^{-1} d\tilde z_r
\end{eqnarray}
where we have introduced $h_\eta(z)=i \delta(z-1)$, while $h_{\chi}(z)=h(z^{-1})$.
We now note that for each propagator, we contribute a factor of $\beta^{-1}$. A rapid calculation will show that for each $h_{\chi}$ there is a factor of $\beta$ and for each vertex there are two lines $\chi$. If $k=0$, since only $\chi$ fields can be contracted, we have a factor of $\beta^{-n}$. For each external $\chi$ field, we have two external factors $\beta$  and one due to losing one internal propagators. This implies that, since $k$ must be even, we have that 
	
We see then that in order to write an expansion in powers of $\beta^{-q}$, we can order them according to:
\begin{equation}
-n+\frac{3}{2}k=-q \rightarrow (n,2k)=\left(3p+q,2k \right)
\end{equation}
with $2p\leq n$, from which we can derive the table for the pair $(n,2k)$ at each perturbation order in $\beta^{-1}=T$.

\begin{table}
\centering
\begin{tabular}{|c|c|c|c|c|c|}
\hline
$\beta^{-q}$ & $k=0$ & $k=1$ & $k=2$ & $k=3$ & $\cdots$ \\
\hline
$\beta^{-1}$ & $(1,0)$ &  $(4,1)$ & $(7,4)$ & $(11,6)$ & $\cdots$ \\ 
\hline
$\beta^{-2}$ & $(2,0)$ & $(5,2)$ &  $(8,4)$ & $(12,6)$  & $\cdots$ \\ 
\hline
$\beta^{-3}$ & $(3,0)$ & $(6,2)$ &  $(9,4)$ & $(13,6)$  & $\cdots$ \\ 
\hline
$\beta^{-4}$ & $(4,0)$ & $(7,2)$ &  $(10,4)$ & $(14,6)$  & $\cdots$ \\ 
\hline
$\vdots$ & $\vdots$ & $\vdots$ & $\vdots$ & $\vdots$ & $\vdots$ \\
\hline
\end{tabular}
\caption{Diagrams contributing to the same order in $\beta^{-1}$. We see that for each order, an infinity of diagrams contribute if $h\neq0$.} 
\label{fig:tablepert}
\end{table}

In Tab. \ref{fig:tablepert} we show that if an external field $h_\chi$ is present, an infinite tower of diagrams contribute to the same order in perturbation theory.
This implies that this perturbation theory is well defined only for $h=0$, in which the number of interaction vertices equals the number of external legs. Thus, in order for the perturbation to make sense, either one sets $h_\chi=0$ or performs a background field expansion. Let us thus consider first $h_\chi=0$. 
\begin{eqnarray}
 W[h_\eta]=\sum_{n=0}^\infty \frac{(4\pi^2)^n}{\beta^n n!} \oint_{\mathcal C}dz_1 &\cdots& dz_n  G^{n}_{conn}(z_1,\cdots,z_n) \nonumber \\
&\cdot& h_{\eta}(z_1)\cdots h_{\eta}(z_n)
\end{eqnarray}
It is easy to see that $\mathbb E[ \chi(z)]=0$ if $h_\chi=0$.  If also $J=0$, which is equivalent to the tree level, then the only diagram which is viable is the vertex, and also the only diagram which contributes at the zeroth order. We have
\begin{eqnarray}
\mathbb E[ \chi(z_1)\chi(z_2)]&=&(2 \pi i)^2\frac{i}{4\pi^2} \oint dz\ z^{-1} \delta(z-z_1^{-1} z_2^{-1} )\delta(z-1)\nonumber \\
&=&  \delta(1-z_1^{-1} z_2^{-1})
\end{eqnarray}
Thus
\begin{equation}
\mathbb E[ \sigma_i \sigma_j ]= -\frac{1}{4\pi^2}\oint dz_1 z_1^i dz_2 z_2^{j} \mathbb E[ \chi(z_1) \chi(z_2) ]=\delta_{ij}
\end{equation}
which is what one would expect. Thus, the argument that brought us to the field theoretical approach is recovered.
At the second order,  the first diagram which contributes to $\beta^{-1}$ has $n=1$. The first average is thus
\begin{eqnarray}
\mathbb E[ \chi(\tilde z_1)\chi(\tilde z_2)]&=&\sum_{n=2}^\infty  \frac{(4\pi^2)^{n-1}}{\beta^{n-1} n!} \oint_{\mathcal C}  G_{\text{conn}}^n(\tilde z_1, z_2,\tilde z_1,\cdots, z_{n}) \nonumber \\
&\cdot&\prod_{i=1}^n d z_i z_i^{-1}J_{\eta}( z_i) 
\end{eqnarray}
\begin{figure}
\centering
\includegraphics[scale=0.2]{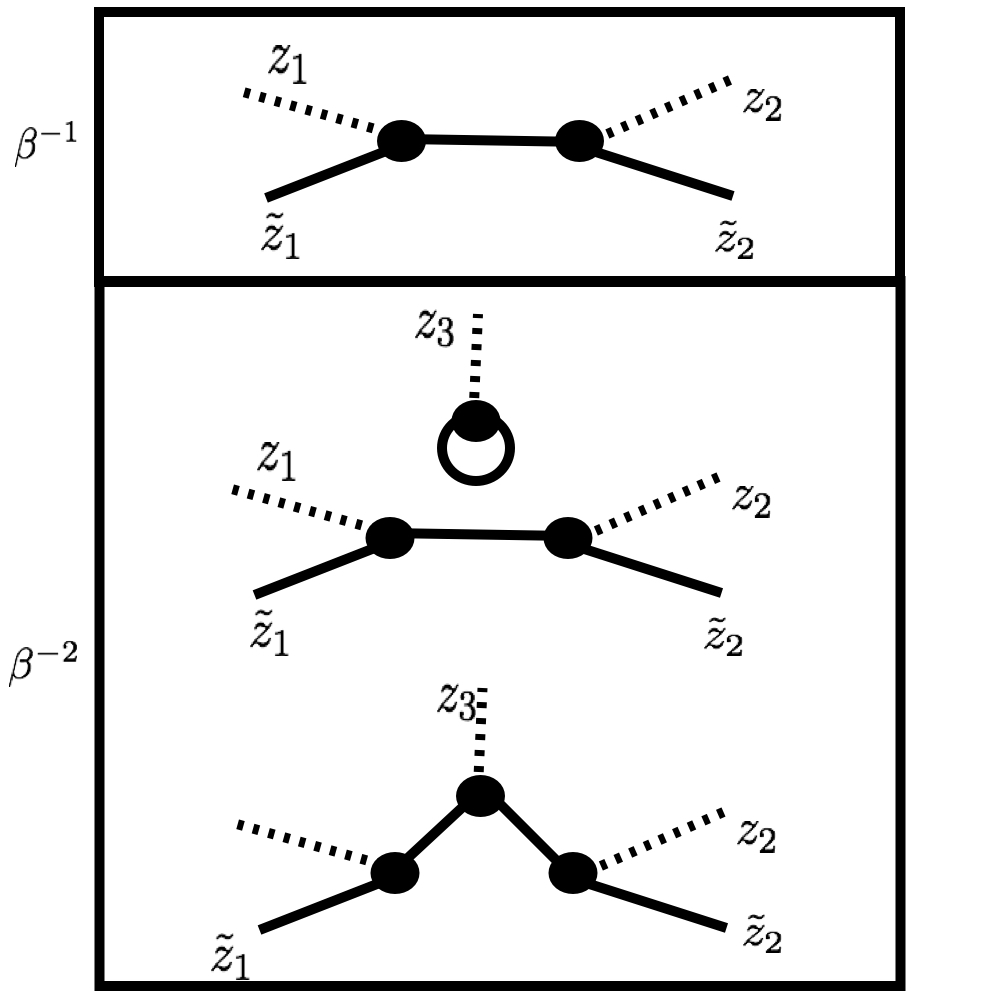}
\caption{Diagrams contributing to $Z \mathbb E[ \chi(z_1) \chi(z_2) ]$ up to order $\tilde \beta^{-2}$ in the $m\rightarrow 0$ limit.}
\label{fig:chichi}
\end{figure}
We consider the diagrams in Fig. \ref{fig:chichi}. The first order term gives the following integral to evaluate:
\begin{eqnarray}
 &2&\frac{1}{(4\pi^2)^2} \oint_{\mathcal C}\oint_{\mathcal C}dz_1 dz_2 dz_a dz_b z_a^{-1} z_b^{-1}z_1^{-1} dz_2^{-1} \nonumber \\
 &\cdot& \tilde J^{-1}(z_a^{-1},z_b^{-1}) (-1)^2 \delta(z_1-z_a \tilde z_1^{-1})\delta(z_2-z_2 \tilde z_2^{-1}) \nonumber \\
 &=&\frac{2}{(4\pi^2)^2} \tilde J^{-1}(\tilde z_1^{-1},\tilde z_2^{-2})
\end{eqnarray}
From which we obtain
\begin{eqnarray}
\mathbb E[ \sigma_i \sigma_j ]&=&\frac{(4\pi^2)^2}{2 \beta}\oint d\tilde z_1 d\tilde z_2 \tilde z_1^{-1} \tilde z_2^{-2} \mathbb E[ \chi(\tilde z_1) \chi(\tilde z_2) ] \nonumber \\
&=&\delta_{ij}+T J_{ij}^{-1}+O(T^2)
\end{eqnarray}
The first term above is the same that we would obtain if we had a partition function of the form:
\begin{equation}
Z(\vec h)=\int d\sigma_i e^{ -\beta \left(\sum_{ij} \sigma_i \sigma_j J_{ij} +\sum h_i \sigma_i\right)}
\end{equation}
for continuous ``spin" variables $\sigma_i$, and evaluated the covariance at zero external field. However, since we have spin variables, we obtain higher order corrections (in temperature).
The second order has two contributing diagrams. 
The first diagram gives is
\begin{equation}
\frac{2}{(4 \pi^2)^3} \text{Trace}(J^{-1}) \tilde J^{-1}(\tilde z_1^{-1},\tilde z_2^{-1})
\end{equation}
while the second gives
\begin{equation}
\frac{2}{(4 \pi^2)^3}  \tilde J^{-2}(\tilde z_1^{-1},\tilde z_2^{-1})
\end{equation}
and thus we obtain, at the second order
\begin{equation}
\mathbb E[ \sigma_i \sigma_j ]=\delta_{ij}+T J_{ij}^{-1}+\frac{T^2}{3} (\text{Trace}(J^{-1}) J_{ij}^{-1}+ J_{ij}^{-2})+O(T^3)
\end{equation}
At the order $T^q$, the terms that appear are of the form
\begin{equation}
T^q:\ \  J^{-m}_{ij}\prod_{i=1}^f \text{Trace}(J^{-r_i})
\end{equation}
with $\sum_{i=1}^f r_i+m=q$, where $f$ is the number of closed loops in each diagram,  and  $r_i$ is the length of each loop. An open string going from one incoming to outcoming line of length $m$. 
The symmetry coefficient in front of each term is equal to the number of ways we can construct that diagram. These features generalize also to $n$-point functions.

The coefficients can be evaluated as follows. 
At each order $q$ there are $q$ vertices. Let us assume that there are two $\chi$ external lines. Then the length of the open line is parametrized as
$k=1,\cdots,q-1$. Thus, at order $q$, we can thus parametrize the correction as
\begin{equation}
\tilde T^q_2:\ \frac{1}{q!}\sum_{k=1}^{2(q-1)} D_k^q J^{-k}_{ij}R_k
\end{equation}
where $D_k$ is the combinatorial factor associated with the number of ways we can draw an open line of length $k$. This factor is easy to see that it is the number of possible ways of choosing the $k$ nodes out of $n$, which is $D_k^q=\frac{q!}{(k-1)!}$.
The factor $R_k$ is decomposed as we have seen above, with $\prod_{i=1}^f \text{Trace}(J^{-r_i})$, with a combinatorial factor which can be thought in the number of ways we can partition the remaining $n-k$ nodes. The factors $r_i$ are thus connected to the problem of partitioning a set of $q-k$ elements.
The total number of partitions is given by the Bell numbers, $B_{q-k}$, but we are interested, exactly, in the number of partitions with the same grouping. This is given by the Stirling numbers of the second kind.
For instance, the total number of partitions of a set of $q-k=5$ elements is 52.
However, there is only one way of grouping all nodes together, which results in $\text{Trace}(J^{-5})$, and one way of having all nodes grouped alone, which results in $\text{Trace}(J^{-1})^5$. Also, we need to consider the combinatorial factors associated with each loop, and an overall factor $\frac{1}{2}$ associated with the automorphism of the open line. There are however more, and thus we have 
\begin{itemize}
\item 15 ways as $\{a,\{b,c\},\{d,e\}\}$, which results in $\text{Tr}(J^{-1})\text{Tr}(J^{-2})^2$, and a factor $\frac{1}{4}\cdot\frac{1}{4}\cdot 4\cdot 4=1$ ;
\item 10 ways as $\{a,b,c,\{d,e\}\}$, equivalent to $\text{Tr}(J^{-1})^3\text{Tr}(J^{-2})$, and a factor $\frac{1}{4}\cdot 4=1$;
\item 10 ways as $\{\{a,b\},\{c,d,e\}\}$, equivalent to $\text{Tr}(J^{-2})\text{Tr}(J^{-3})$ and a factor $\frac{1}{4}\cdot \frac{1}{6}\cdot 4 \cdot 24=4$;
\item 10 ways as $\{a,b,\{c,d,e\}\}$ which corresponds to $\text{Tr}(J^{-1})^2\text{Tr}(J^{-3})$ and a factor $\frac{1}{6} \cdot 24=4$;
\item 5 ways as $\{a,\{b,c,d,e\}\}$ equivalent to $\text{Tr}(J^{-1})\text{Trace}(J^{-4})$ and a factor $\frac{1}{8}*1920=24$. 
\item 1 way as  $\{\{a,b,c,d,e\}\}$ equivalent to $\text{Tr}(J^{-5})$ and a factor $\frac{1}{10} \cdot 1920=192$. 
\item 1 way as  $\{\{a\},\{b\},\{c\},\{d\},\{e\}\}$ equivalent to $\text{Tr}(J^{-1})^5$ and a factor $1$. 
\end{itemize}
Thus, at the 8th order we have a term of the form
\begin{eqnarray}
&\frac{1}{2}\frac{T^8}{5!}& J_{ij}^{-2}\Big(192\text{Tr}(J^{-5})+\text{Tr}(J^{-1})^5+120 \text{Tr}(J^{-1})\text{Tr}(J^{-4})+10\text{Tr}(J^{-1})^3\text{Tr}(J^{-2})  \nonumber \\
&+&40\text{Tr}(J^{-2})\text{Tr}(J^{-3})+40 \text{Tr}(J^{-1})^2\text{Tr}(J^{-3})+15\text{Tr}(J^{-1})\text{Tr}(J^{-2})^2 \Big) 
\end{eqnarray}
and, analogously, also a term
\begin{eqnarray}
5 T^8 J_{ij}^{-6} \text{Tr}(J^{-1}),
\end{eqnarray}
among others at the same order.

\section{A few final remarks and future work}
In this paper we introduced a formalism to study partition functions of discrete models, such as the Ising model, as a statistical field theory over the complex plane. From a field theoretical point of view, the Ising model can be represented as scalar Yukawa model in which the temperature becomes the coupling of the interaction, and the quadratic term typical of the Ising model the propagator for the ``Ising" field. The spin-like nature of the spins are introduced via an auxiliary field.
This implies a fractionalization of the Ising field, and the Landau-Ginzburg model (quartic interaction) is recovered after the integration of one field, interpreting the auxiliary field as a intermediate field representation.
We have seen that from that a field theoretic (diagrammatic) point of view, the Ising model can be obtained from the ``urchin diagram" expansion: only diagrams with external line of one field have to be considered. The formalism 
Here we further comment that what we have developed is a specific type of Group Field Theory over the $U(1)$ group Haar measure.
In order to see this, let us define
\begin{eqnarray}
Z&=&\int [D\chi(z)] [D\eta(z)]e^{-\oint_{\mathcal C} dz z^{-1} \eta(z) \delta(z-1)} \nonumber \\
&\cdot& e^{ \frac{1}{4 \pi^2 } \oint_{\mathcal C}\oint_{\mathcal C} dz_1 dz_2 \chi(z_1)\chi(z_2)z_1 ^{-1} z_2^{-2}\left(-i\eta(z_1^{-1} z_2^{-1})+ \beta J(z_1^{-1},z_2^{-1})\right)}\nonumber \\
&\cdot&e^{-i\beta \frac{1}{2\pi } \oint_{\mathcal C} dz_1 z_1^{-1} \chi(z_1) h(z_1^{-1}) }
\label{eq:fieldtheory2}
\end{eqnarray}
as a group field theory. When $\mathcal C$ is the unit circle, we have that $z=e^{i\theta}$ which can be identified as an element of the $U(1)$ group \cite{Oriti}. Specifically, we can write

\begin{eqnarray}
Z=\int [D\chi(g)] [D\eta(g)]& &e^{-2\pi \int dg g^{-1}\ \eta(g) \delta(g)-\beta \int dg_1dg_2 g_1^{-1}g_2^{-1}  \chi(g_1)\chi(g_2) J(g_1^{-1},g_2^{-1}) }  \nonumber \\
&\cdot& e^{-i2\pi \int dg_1 dg_2 dg_3 g_1^{-1}g_2^{-1}  \chi(g_1)\chi(g_2)\eta(g_3) \delta(g_1 g_2 g_3)-i\beta \int dg g^{-1}\chi(g) h(g^{-1}) }
\label{eq:gfieldtheory}
\end{eqnarray}
where $\int dg$ is the Haar measure over the $U(1)$ group and $\delta(g)$ is the delta function over the $U(1)$ group. The model above can be then interpreted as a Group Field Theory, which is a field theory where the base manifold is a group.
 Interestingly, the conservation of momentum valid for the Fourier transform
when the model is translationally invariant, is valid in the $U(1)$ formulation also when translation invariance is not present.
We believe that the model we have introduced is useful for a few reasons. The representation via the complex integrals can be a useful tool, as we have shown, to map discrete to continuous fields without a continuous limit. We have nonetheless shown that the discreteness of is contained in the propagator and in the auxiliary field interaction.
We have also noted in the resummation of the exact series in the temperature-rescaling of the auxliary field mass that the perturbative ``pole" which we introduced by hand in $z=0$, moves to $z=1$. This is an interesting feature which we have observed in the analysis performed in this paper.

Another comment we would like to make is that this transformation can be performed for arbitrary tensor fields.
Consider for instance the following $p-$spin interaction:
\begin{equation}
    Z=\sum_{\{\sigma_i=\pm 1\}} e^{\beta \sum_{i_1 \cdots i_k}J_{i_1\cdots i_k} \sigma_{i_1}\cdots \sigma_{i_k}}.
\end{equation}
If we introduce the z-Transform as for the Ising model, we obtain
\begin{equation}
    Z=\int [D\eta][D\chi] e^{ \frac{\beta }{(2\pi i)^k } \oint \frac{dz_1 \cdots dz_k}{z_1 \cdots z_k} J(z_1^{-1},\cdots,z_k^{-1}) \chi(z_1)\cdots \chi(z_k) + \frac{1}{(2\pi)^2 i} \oint \frac{dz}{z} \eta(z_1 z_2) \chi(z_1^{-1}) \chi(z_2^{-2}) -\oint dz \eta(z) \delta(z)},
\end{equation}
with an obvious definition for $J(z_1,\cdots,z_k)$, and thus in principle such formalism can be used for more complicated type of interactions.

It is important to mention that in the present paper we have inserted by hand the mass for the auxiliary field. However, for the Ising model the ``mass``, or quadratic term, can be introduced via the expansion at the second order in a Loop-Vertex expansion, interpreting the field theory as an intermediate field representation \cite{rivasseau}. This will be the focus of a future work \cite{inprep}. It is also worth mentioning that the z-Transform technique can be used for  discrete models (for instance, tensor models\cite{gurau}) in which one has contractions between discrete indices. Also, in future works we will be interested in studying quenched averages of the free energy using constructive field theory methods.\ \\\ \\

\textbf{Acknowledgements}{
This work was carried out under the auspices of the NNSA of the U.S. DoE at LANL under Contract No. DE-AC52-06NA25396. FC was also financed via DOE-ER grants PRD20170660 and PRD20190195. 

I would like to thank several people. First, C. Nisoli, V. Cirigliano and A. Saxena for encouragement to work on this topic, and D. Oriti and A. Trombettoni for comments in the early stages of this work. Also, several important remarks were made by R. Gurau and S. Carrozza during a visit at Perimeter Institute, which will be developed in a follow up to this paper. I also would like to thank D. Facoetti for some elucidating discussions at Parisi70 in Rome, on the possibility to use this approach to study glassy systems, and C. Castelnovo for discussions during his visit at LANL.

At last, I had the idea for this paper during an X-Ray at S. Vincent Hospital in Santa Fe. I would like to thank the doctor for a joke he made that sparked the idea, as we were discussing the z-Transform.

\textit{The pilots of an airplane are left unconscious in an air pressure bump. The hostess asks if anybody can fly the airplane. A mathematician raises his hand saying he has an idea of how to do it. Put in front of the cloche, the hostess shouts ``move! we need to get the airplane stable as soon as possible". The mathematician responds: ``Calm down, it's only a simple pole in a complex plane".}

}

\appendix

\section{Lattice Green functions}
We now consider the coupling matrix $J_{ij}$ on lattices and their integral representation in $D$ dimensions.
We consider a coupling which has a $D$-dimensional discrete Euclidean symmetry.
\begin{equation}
J_{ij}=
\begin{cases}
A \text{ if } i=j \\
-B \text{ if } i,j \text{ neighbors} 
\end{cases}
\end{equation}
and describes a hyperdimensional lattice. The natural representation for the matrix $J_{ij}$ is the Fourier transform, defined on the vectors $\vec r_i=(x^1_i,x^2_i,\cdots,x^D_i)$ which defines the location of the particle on the lattice. This is given by
\begin{equation}
J_{km}=\int_{-\pi}^\pi \frac{d^Dp}{(2\pi)^D} G(\vec p) e^{i(\vec r_k-\vec r_m)\cdot \vec p}.
\end{equation}
The inverse (if it exists) is given by
\begin{equation}
J_{km}^{-1}=\int_{-\pi}^\pi \frac{d^Dp}{(2\pi)^D} G(\vec p)^{-1} e^{i(\vec r_k-\vec r_m)\cdot \vec p}.
\end{equation}
Let us assume that $J_{km}=\alpha I+ K A_{ij}$ where $A_{ij}$ is the adjacency matrix of a lattice. Then
\begin{equation}
G(\vec p)=\alpha+ 2 K \sum_{\nu=1}^D \cos(p_\nu).
\end{equation}
The formula above is useful to evaluate $J(z_1,z_2)$ and its Green function $J^{-1}(z_1,z_2)$.

{
\section{Partition function with temperature rescaling up to the $6th$ order }
We report below the first 7 terms of the low temperature expansion of partition function, found using a computer software written in Matlab to evaluate all the Wick contractions up to $T^6$ in the low temperature expansion. The result is:
\begin{eqnarray}
\sqrt{T}\frac{Z_{Ising}}{Z_G}&=&1+T\ \text{Tr}\left(J^{-1}\right)+\frac{T^2}{2!}\Big(\left(\text{Tr}\left(J^{-1}\right)\right)^{2}+2\text{Tr}\left(J^{-2}\right)\Big) \nonumber \\
&+&\frac{T^3}{3!}\Big(\left(\text{Tr}\left(J^{-1}\right)\right)^{3} +6\text{Tr}\left(J^{-1}\right)\text{Tr}\left(J^{-2}\right)+8\text{Tr}\left(J^{-3}\right)\Big) \nonumber \\
&+&\frac{T^4}{4!}\Big(\left(\text{Tr}\left(J^{-1}\right)\right)^{4}
+12\left(\text{Tr}\left(J^{-1}\right)\right)^{2}\text{Tr}\left(J^{-2}\right)+32\text{Tr}\left(J^{-1}\right)\text{Tr}\left(J^{-3}\right)+12\left(\text{Tr}\left(J^{-2}\right)\right)^{2} 
+48\text{Tr}\left(J^{-4}\right)\Big)\nonumber \\
&+&\frac{T^5}{5!}\Big(\left(\text{Tr}\left(J^{-1}\right)\right)^{5}+20\left(\text{Tr}\left(J^{-1}\right)\right)^{3}\text{Tr}\left(J^{-2}\right) \nonumber \\
&+&80\left(\text{Tr}\left(J^{-1}\right)\right)^{2}\text{Tr}\left(J^{-3}\right) 
+60\text{Tr}\left(J^{-1}\right)\left(\text{Tr}\left(J^{-2}\right)\right)^{2}+240\text{Tr}\left(J^{-1}\right)\text{Tr}\left(J^{-4}\right) \nonumber \\
&+&160\text{Tr}\left(J^{-2}\right)\text{Tr}\left(J^{-3}\right)+384\text{Tr}\left(J^{-5}\right)\Big) \nonumber \\
&+&\frac{T^6}{6!}\Big(\left(\text{Tr}\left(J^{-1}\right)\right)^{6}+30\left(\text{Tr}\left(J^{-1}\right)\right)^{4}\text{Tr}\left(J^{-2}\right)\nonumber \\
&+&160\left(\text{Tr}\left(J^{-1}\right)\right)^{3}\text{Tr}\left(J^{-3}\right)
+180\left(\text{Tr}\left(J^{-1}\right)\right)^{2}\left(\text{Tr}\left(J^{-2}\right)\right)^{2}\nonumber \\
&+&720\left(\text{Tr}\left(J^{-1}\right)\right)^{2}\text{Tr}\left(J^{-4}\right)\nonumber \\
&+&960\text{Tr}\left(J^{-1}\right)\text{Tr}\left(J^{-2}\right)\text{Tr}\left(J^{-3}\right)\nonumber \\
&+&2304\text{Tr}\left(J^{-1}\right)\text{Tr}\left(J^{-5}\right)\nonumber \\
&+&120\left(\text{Tr}\left(J^{-2}\right)\right)^{3}+1440\text{Tr}\left(J^{-2}\right)\text{Tr}\left(J^{-4}\right)\nonumber \\
&+&640\left(\text{Tr}\left(J^{-3}\right)\right)^{2}+3840\text{Tr}\left(J^{-6}\right)\Big)
\end{eqnarray}
where $Z_G$ is the Gaussian model partition function.
}

\section{Variation of functional and ``Cauchy" delta functions}
In this section we comment on the variation of an integral of the form:
\begin{equation}
F_n[\chi(z)]=\frac{1}{(2\pi i)^n} \oint_{\mathcal C} \delta( \prod_{i=1}^n z_i-1)\prod_{i=1}^n\chi(z_i) \frac{dz_i}{z_i}.
\end{equation}
The variation is defined as 
\begin{equation}
\frac{\delta}{\delta \chi(z_0)} F_n[\chi(z)]=\frac{d}{d\epsilon} F_n[\chi(z)+\epsilon \delta(z-z_0)]
\end{equation}
where the Delta function is intended as the Cauchy delta function. It is easy to see that:
\begin{eqnarray}
\frac{\delta^k}{\delta \chi(\tilde z_1)\cdot \delta \chi(\tilde z_k)} F_n[\chi(z)]&=&\frac{n!}{(n-k)! (2\pi i)^n \prod_{i=1}^k \tilde z_i} \cdot \nonumber \\
&\cdot&\oint_{\mathcal C}\delta( \prod_{i=1}^n z_i-1)\prod_{i=k+1}^n\chi(z_i) \frac{dz_i}{z_i}. \nonumber \\
\end{eqnarray}
Something that it is useful is the case for ``constant fields". In this case $\chi(z)\rightarrow -2\pi i \langle \chi \rangle \delta(z-1)$, and one has
\begin{equation}
\frac{\delta^k}{\delta \chi(\tilde z_1)\cdot \delta \chi(\tilde z_k)} F_n[\chi(z)]=\frac{n! (-1)^{n-k}}{(n-k)! (2\pi i)^{k}} \langle \chi \rangle^{n-k} \delta(0)
\end{equation}
and from which we obtain
\begin{equation}
Tr\left(\frac{\delta^2}{\delta \chi(\tilde z_1) \delta \chi(\tilde z_2)} F_n[\chi(z)]\right)=\frac{n(n-1) (-1)^{n-2}}{(2\pi i)^{2}} \langle \chi \rangle^{n-2}  \delta(0)
\end{equation}
which is the formula used in the main text.

\section{Representation of the delta over the U(1) group}
Let introduce the following Haar measure:
\begin{equation}
 \int_{-\pi}^{\pi} \frac{d \theta}{2 \pi}=1
\end{equation}
We want to show that we can represent the delta over the U(1) group in the following way:
\begin{equation}
 \delta(g)\equiv \delta(\theta)=\sum_{n=-\infty}^{\infty} e^{i n \theta}
\end{equation}
Let note that $e^{i n \theta}$ is a complete orthonormal basis for the Hilbert space on the unit circle with the scalar product given by:
$$<f,g>= \int_{-\pi}^{\pi} \frac{d \theta}{2 \pi} f^{*}(\theta) g(\theta)$$
We have to show the following properties for the $\delta$:
\begin{equation}
  \int_{-\pi}^{\pi} \frac{d \theta}{2 \pi} \delta(\theta) = 1
\label{prop1}
\end{equation}
\begin{equation}
  \int_{-\pi}^{\pi} \frac{d \theta}{2 \pi} \delta(\theta) f(\theta) = f(0)
\label{prop2}
\end{equation}
Equations (\ref{prop1}) and (\ref{prop2}) follow from the orthogonality of the vectors:
\begin{equation}
 <e^{i n_1 \theta},e^{i n_2 \theta}>=\int_{-\pi}^{\pi} \frac{d \theta}{2 \pi} e^{i (n_2-n_1)\theta}=\delta_{n1,n2}
\end{equation}
Property (\ref{prop1}) follows from:
\begin{equation}
 <1,\delta(\theta)>=\int_{-\pi}^{\pi} \frac{d \theta}{2 \pi} \sum_{n=-\infty}^{\infty} e^{i n \theta}=\sum_{n=-\infty}^{\infty} \delta_{n,0}=1
\end{equation}
Now every function over the circle has the following Fourier decomposition:
$$f(\theta)=\sum_{n=-\infty}^{\infty} c_n e^{in \theta}$$
and:
$$f(0)=\sum_{n=-\infty}^{\infty} c_n $$
So property (\ref{prop2}) comes from:
\begin{eqnarray}
& <f^*(\theta),\delta(\theta)>=\int_{-\pi}^{\pi} \frac{d \theta}{2 \pi} \sum_{n_1=-\infty}^{\infty} c_{n_1} e^{i n_1 \theta} \sum_{n_2=-\infty}^{\infty} e^{i n_2 \theta}=\nonumber \\
& = \sum_{n_1,n_2} c_{n_1} \delta_{n_1,-n_2}=\sum_{n_1} c_{n_1}=f(0) \\
\end{eqnarray}
and thus we have a good representation of the delta over the circle. \\\ \\

\end{document}